\documentclass[acmsmall]{acmart}
\usepackage{enumitem}
\usepackage{multirow}
\usepackage{bm}
\usepackage{subfigure}
\usepackage{makecell}
\usepackage{tcolorbox}
\usepackage{soul}
\usepackage{amsmath}
\usepackage{ulem}
\usepackage{hyperref}
\usepackage{cleveref}
\usepackage{hyperxmp}
\usepackage{url}
\usepackage{graphicx}

\newcommand\etal{{\it{et al.\ }}}

\usepackage{xcolor}

\usepackage{listings}

\definecolor{solidityblue}{RGB}{46, 107, 186}
\definecolor{soliditygreen}{RGB}{41, 153, 30}
\definecolor{soliditygrey}{RGB}{128, 128, 128}
\definecolor{soliditypurple}{RGB}{155, 66, 237}

\lstdefinelanguage{Solidity}{
    keywords={pragma, solidity, contract, library, interface, is, function, modifier, event, public, private, external, internal, returns, return, emit, mapping, address, uint256, bool, string, memory, storage, calldata, import, constructor, require, if, else, for, while},
    keywordstyle=\color{solidityblue}\bfseries,
    ndkeywords={Ownable, ReentrancyGuard, IERC20}, 
    ndkeywordstyle=\color{soliditypurple}\bfseries,
    identifierstyle=\color{black},
    sensitive=false,
    comment=[l]{//},
    morecomment=[s]{/*}{*/},
    commentstyle=\color{soliditygrey}\itshape,
    stringstyle=\color{soliditygreen}\ttfamily,
    morestring=[b]",
    morestring=[b]'
}

\AtBeginDocument{%
  }

\setcopyright{acmlicensed}
\copyrightyear{2018}
\acmYear{2018}
\acmDOI{XXXXXXX.XXXXXXX}

\acmJournal{JACM}
\acmVolume{37}
\acmNumber{4}
\acmArticle{1}
\acmMonth{8}

\begin{document}

\title{Repository-Level Solidity Code Generation with Large Language Models: From Prompting to Fine-Tuning}

\author{Shi Chen}
\email{ts24170102p31@cumt.edu.cn}

\author{Rongcun Wang}
\authornote{Corresponding author.}
\email{rcwang@cumt.edu.cn}
\affiliation{
  \institution{School of Computer Science and Technology (School of Artificial Intelligence), China University of Mining and Technology}
  \city{Xuzhou}
  \state{Jiangsu Province}
  \country{China}
  \postcode{221116}
}
\affiliation{
  \institution{Mine Digitization Engineering Research Center of the Ministry of Education}
  \city{Xuzhou}
  \state{Jiangsu Province}
  \country{China}
  \postcode{221116}
}
\affiliation{
  \institution{Jiangsu Provincial Industrial Technology Engineering Center for Intelligent Sensing and Emergency IoT in Underground Space}
  \city{Xuzhou}
  \state{Jiangsu Province}
  \country{China}
  \postcode{221116}
}

\author{Yuan Tian}
\email{y.tian@queensu.ca}
\affiliation{
  \institution{School of Computing, Queen’s University}
  \city{Kingston}
  \country{Canada}
}

\author{Xiaoyuan Xie}
\email{xxie@whu.edu.cn}
\affiliation{
  \institution{School of Computer Science, Wuhan University}
  \city{Wuhan}
  \state{Hubei Province}
  \country{China}
    \postcode{430072}
}

\author{Wei Song}
\email{wsong@njust.edu.cn}
\affiliation{
  \institution{School of Computer Science and Engineering, Nanjing University of Science\&Technology}
  \city{Nanjing}
  \state{Jiangsu Province}
  \country{China}
  \postcode{210094}
}
\author{Rubing Huang}
\email{rbhuang@must.edu.mo}
\affiliation{
  \institution{School of Computer Science and Engineering, Macau University of Science and Technology}
  \city{Macau}
  \country{China}
  \postcode{999078}
}


\renewcommand{\shortauthors}{S. Chen et al.}

\begin{abstract}
    
Large Language Models (LLMs) have exhibited remarkable capabilities in general-purpose code generation. However, achieving comparable performance in specialized software domains remains challenging, since generic pre-training alone often fails to capture domain-specific patterns. Smart contracts represent a high-impact domain for studying domain-specific code generation, as they are extensively used in practice, difficult to remediate after deployment, and frequently associated with high-stakes financial assets. Written in domain-specific languages such as Solidity, smart contracts impose specialized constraints at both the language and software levels. Despite these elevated demands on code quality, the performance and domain-specific challenges of Solidity code generation remain underexplored, largely due to the lack of comprehensive benchmarking datasets and evaluation metrics specifically designed to assess the semantic correctness of generated Solidity code. To fill this gap, we introduce a new benchmark consisting of 5,470 high-quality, repository-level Solidity smart contracts paired with natural language descriptions. To our knowledge, this dataset is the first of its kind in terms of scale and quality for the systematic evaluation of Solidity code generation. In addition, we propose \textit{SolidityScore}, a semantics-aware evaluation metric that prioritizes domain-critical Solidity constructs, such as security modifiers, over surface-level token matching. Leveraging on this benchmarking framework, we conduct an empirical evaluation of representative code LLMs, including Qwen2.5-Coder, DeepSeek-Coder, and CodeLlama, across multiple adaptation paradigms, including zero-shot prompting, Chain-of-Thought (CoT) reasoning, in-context learning (ICL), retrieval-augmented generation (RAG), and supervised fine-tuning (SFT). Our results show that: (1) general-purpose models exhibit systematic structural deficiencies, particularly in handling Solidity-specific constructs; (2) among non-parametric methods (that do not update model parameters), RAG achieves the strongest performance, while ICL suffers from rapid context saturation beyond two examples; and (3) SFT emerges as the most effective adaptation strategy, yielding substantial improvements in semantic correctness by internalizing Solidity-specific constraints directly into model parameters. Overall, our work establishes a comprehensive benchmark for repository-level Solidity code generation and identifies the combination of high-quality domain data and SFT as the most effective strategy for improving the reliability of LLM-generated smart contracts.

\end{abstract}





\maketitle

\section{Introduction}
\label{chap:chap1}

The rapid advancement of Large Language Models (LLMs) has begun to reshape software engineering. Both general-purpose (e.g., GPT-4~\cite{GPT42023}) and code-oriented foundation models (e.g., CodeLlama~\cite{DBLP:journals/corr/CodeLlama}, DeepSeek-Coder~\cite{DBLP:journals/DeepSeek-Coder}), have demonstrated strong capabilities in core software engineering tasks, including automated code completion~\cite{intro:cc/Bairi24,intro:cc/Izadi24,intro:cc/Liu24}, code generation~\cite{Tambon2025,Jiang2025}, code translation~\cite{Macedo00CA25}, and program repair~\cite{intro:pr/Chen24,intro:pr/Hossain24,intro:pr/Yin24}. However, reported successes with LLM-based code generation are largely concentrated in popular general-purpose programming languages (GPLs), such as Python, due to the design and scope of current evaluation benchmarks. Widely adopted benchmarks such as HumanEval~\cite{DBLP:journals/CODEX}, MBPP~\cite{DBLP:journals/MBPP}, BigCodeBench~\cite{DBLP:conf/BigCodeBench}, and TACO~\cite{DBLP:journals/TACO} primarily target standalone functions and general algorithmic reasoning in GPLs. As a result, the behavior, limitations, and failure modes of LLMs in domain-specific software and domain-specific programming languages (DSLs) remain underexplored. This gap is critical, as such languages are typically underrepresented in general-purpose pre-training corpora and require specialized domain knowledge, conditions under which LLMs are prone to hallucinations, incomplete semantic understanding, and misuse of critical APIs. In practice, general-purpose LLMs frequently struggle to produce reliable repository-level Solidity code due to a lack of domain-specific knowledge. Generated outputs are often plagued by syntax errors, rendering them uncompilable. More critically, even when the generated code is syntactically valid, it often harbors high-risk, domain-specific security vulnerabilities—such as reentrancy attacks~\cite{DBLP:journals/PoCo}, which can lead to catastrophic financial losses. These limitations highlight the urgent need for a systematic investigation into domain-specific code generation.

In this work, we focus on repository-level Solidity code generation for smart contracts, an emerging yet high-stakes domain of software development. Solidity, the core programming language of the Ethereum ecosystem, poses unique challenges that distinguish it sharply from general-purpose languages. It combines object-abstractions with multi-dimensional constraints, including gas consumption models, event-driven execution semantics, and strict security patterns \cite{intro:journals/Optimal,intro:proceeding/GASOL,intro:conf/proceeding/Silence}. Moreover, due to the immutability of deployed smart contracts and the substantial financial value they often manage, correctness and reliability are not merely performance considerations but fundamental security requirements. While recent benchmarks have attempted to address Solidity-specific evaluation, they remain insufficient for repository-level code generation. BenchSol \cite{intro:Benchsol/Daspe24}, for example, contains only 15 manually curated samples, severely limiting its representativeness. Even though SolEval~\cite{intro:soleval/Peng25} advances the field by incorporating repository-level context, it primarily operates under a function completion paradigm. By providing the surrounding code structure and dependencies as input, this approach simplifies the repository-level generation challenge. Consequently, existing benchmarks fail to adequately capture the complexity of constructing complete Solidity smart contracts from scratch, particularly regarding global architectural logic, inheritance hierarchies, and complex inter-contract dependencies that characterize real-world Solidity repositories.

Beyond dataset limitations, evaluating repository-level Solidity code generation poses unique challenges for existing evaluation metrics. Unlike typical Python repositories, Solidity projects often exhibit relatively shallow directory hierarchies, with complexity arising instead from their logical topology. Contracts commonly involve inheritance chains, cross-contract calls, shared state variables, strict compiler-version requirements, and reliance on versioned audited external libraries (e.g., OpenZeppelin contracts)~\cite{3.2:Etherscan/Durieux20}. Without reconstructing this precise dependency environment, which is particularly challenging in Solidity due to compiler-version sensitivity, transitive inheritance, and dependence on audited external libraries, generated code is frequently uncompilable, even when its core logic is correct. As a result, execution-based evaluation metrics such as Pass@k~\cite{intro:soleval/Peng25} become overly strict and misaligned with the logical correctness of generated code. On the other hand, widely used static and semantic evaluation metrics face complementary limitations. Surface-level token-matching metrics such as BLEU~\cite{DBLP:conf/acl/BLEU} ignore semantic equivalence, while structural metrics like CodeBLEU~\cite{DBLP:journals/corr/CodeBleu,DBLP:journals/jss/CodeBleu} rely on successful Abstract Syntax Tree construction and are fragile to minor syntax errors. Semantic similarity metrics such as CodeBERTScore~\cite{DBLP:conf/emnlp/CodeBERTScore} depend on encoders pre-trained on general-purpose languages, creating a domain gap that fails to capture Solidity-specific elements such as reserved keywords (e.g., modifier and event), specialized data types (e.g., mapping and address), and contract definition patterns.

In addition to evaluation challenges, the most effective domain adaptation paradigm for repository-level Solidity code generation remains unclear. Although techniques such as Retrieval-Augmented Generation (RAG)~\cite{DBLP:RAG0,DBLP:RAG1,DBLP:RAG2}, Chain-of-Thought reasoning (CoT)~\cite{DBLP:COT0,DBLP:COT1,DBLP:COT2,DBLP:COT3}, In-Context Learning (ICL)~\cite{2.2:icl/Patel24,2.2:icl/MinLZH22,2.2:icl/YangCGLHLX25}, and Supervised Fine-Tuning (SFT)~\cite{DBLP:SFT0,DBLP:SFT1,DBLP:SFT2} have demonstrated promise in other domains, there has been no systematic empirical study that quantitatively compares their relative effectiveness or characterizes the conditions under which they perform well for repository-level Solidity smart contract generation.

To address these gaps, we construct a large-scale benchmark named \textit{SolidityBench}. It comprises 5,470 repository-level Solidity smart contract samples, each paired with a natural language specification that captures repository-level functionality and design intent. The dataset is curated from authoritative sources, i.e., OpenZeppelin\footnote{\url{https://www.openzeppelin.com/}}, Synthetix\footnote{\url{https://synthetix.io/}}, and verified contracts on Etherscan\footnote{\url{https://etherscan.io/}}, and reflects real-world inter-contract dependencies, inheritance structures, and library usage patterns. To address the limitations of existing evaluation metrics, we introduce \textit{SolidityScore}, a domain-aware text-based metric that assesses the semantic alignment between generated code and ground-truth contracts. Unlike execution-based metrics that require a fully reconstructible compilation environment, or surface-level similarity metrics that are insensitive to domain semantics, SolidityScore leverages a Solidity-adapted encoder and domain-weighted token matching to prioritize Solidity constructs. Building on the newly introduced dataset and metric, we conduct a systematic empirical study on representative code LLMs across multiple adaptation paradigms, including CoT, ICL, RAG, and SFT. Specifically, our empirical evaluation is guided by the following research questions:
\vspace{-0.1cm}
\begin{enumerate}[label=\textbf{RQ\textsubscript{\arabic*:}}, leftmargin=3.5em]
    \item How do general-purpose LLMs perform in generating repository-level Solidity code under a zero-shot setting?
    \item To what extent can prompting-based adaptation strategies improve repository-level Solidity code generation?
    \item How does domain-specific supervised fine-tuning compare with prompting-based strategies for repository-level Solidity code generation?
    \item Does \textit{SolidityScore} provide a more reliable assessment of the semantic correctness of generated Solidity code than BLEU?
    \item To what extent can LLMs generate compilable repository-level Solidity contracts, and what are the dominant types of compilation errors?
\end{enumerate}

\vspace{-0.1cm}
Our empirical study reveals that general-purpose LLMs exhibit substantial deficiencies under zero-shot settings; prompting-based strategies provide measurable but limited improvements, with RAG outperforming other non-parametric methods; SFT yields the most significant gains by internalizing Solidity-specific constraints; and compilation remains a major bottleneck, with diverse and recurring error types persisting even in syntactically and semantically plausible outputs.

Our main contributions are summarized as follows:

(1) We curate real-world Solidity repositories from authoritative platforms, including Synthetix, OpenZeppelin, and Etherscan, and construct \textit{SolidityBench}\footnote{\url{https://github.com/ChenS0827/SCG}}, a dataset of 5,470 repository-level natural language specification–code pairs for training and evaluation.

(2) We propose \textit{SolidityScore}, a semantic-aware evaluation metric based on an encoder fine-tuned on Solidity code, enabling more accurate assessment of functional and structural alignment than existing general-purpose metrics.

(3) We establish a systematic benchmark evaluating representative LLMs (e.g., Qwen2.5-Coder, DeepSeek-Coder, and CodeLlama) across multiple adaptation paradigms, including zero-shot inference, CoT, ICL, RAG, and SFT, providing the first comprehensive comparison for repository-level Solidity code generation.

The remainder of this paper is organized as follows. Section \ref{chap:chap2} introduces background and summarizes related work. Section \ref{chap:chap3} presents our benchmarking framework. Section \ref{chap:chap4} details the design and experimental setup of the empirical study. Section \ref{chap:chap5} presents and analyzes the experimental results. Section~\ref{chap:chap6} presents a qualitative case study analyzing the characteristics of all considered Solidity smart contract generation approaches. Section \ref{chap:chap7} discusses implications and threats to validity, followed by the conclusion in Section \ref{chap:chap8}.

\section{Background and Related Work}
\label{chap:chap2}
    
\subsection{Solidity Smart Contract Generation: Challenges and Existing Benchmarks}

Solidity is the core programming language of the Ethereum ecosystem and is used to specify the logic of smart contracts executed on the Ethereum Virtual Machine (EVM). Unlike traditional software systems, Solidity contracts operate in a decentralized and adversarial environment, where deployed code is immutable and often directly manages high-value digital assets, and are executed under a Turing-complete yet resource-constrained execution model~\cite{Nakamoto20227, 2.1：buterin2014next}. These characteristics pose a set of unique requirements and challenges for automated code generation.

First, immutability and adversarial execution impose strict correctness and security requirements. Once deployed, smart contracts cannot be modified, rendering post-deployment fixes costly or infeasible~\cite{2.1:WangCHZBZ23}. In addition, extensive code reuse and cloning practices in the ecosystem amplify the propagation of latent vulnerabilities across projects~\cite{2.1:KhanDVM23}. Many high-risk vulnerabilities—such as reentrancy, improper state updates, and flawed access control—arise from complex interaction patterns rather than superficial syntactic errors~\cite{2.1:GaoJXLG21}. Consequently, automated generation systems must produce secure and correct code at deployment time, without relying on iterative debugging or patching.

Beyond security concerns, Solidity exposes low-level control mechanisms that complicate semantic modeling. To enable fine-grained interaction with the EVM, Solidity allows inline assembly, which appears in approximately 23\% of high-performance contracts~\cite{2.1:ShiXYSG23,2.1:ChaliasosGL22}. Correctly generating such constructs requires an in-depth understanding of EVM execution semantics, storage layout, and calling conventions. However, general-purpose LLMs are primarily trained on high-level programming abstractions, and thus often lack representations of these low-level semantics, leading to logical inconsistencies, reduced readability, or unsafe interactions with EVM internals~\cite{2.1:LiaoSZLHJCCZZ23}.

Finally, Solidity’s gas-based execution model introduces a cost-aware performance dimension absent from most general-purpose languages. Every EVM instruction incurs a monetary cost, and inefficient control flow or suboptimal data structures can result in excessive gas consumption or transaction failure. Prior work has shown that seemingly minor design choices, such as improper loop constructs, may trigger exponential gas growth~\cite{2.1:AlbertCGRR20}. As a result, Solidity code generation must jointly optimize functional correctness, security, and execution efficiency. These objectives are rarely considered simultaneously in existing LLM training or evaluation pipelines.

Despite these unique requirements and challenges, the exploration of LLMs’ potential for Solidity code generation remains at an early stage. To date, only a limited number of benchmarks explicitly target this domain. Among them, SolEval~\cite{intro:soleval/Peng25} is a notable effort that evaluates Solidity code generation while incorporating repository-level context. However, a fundamental distinction lies in its evaluation paradigm compared to a realistic, holistic repository-level generation setting. Specifically, SolEval adopts a function completion approach, in which the LLM is tasked with generating a single function body based on the provided method signatures, requirements, and repository dependencies, which is subsequently integrated back into the codebase for evaluation. This setup substantially simplifies the task by assuming that the contract skeleton and external dependencies are  pre-defined and correct. In contrast, practical development workflows often require \textit{holistic generation}, where developers rely on LLMs to synthesize entire contracts. This demands not only local logic implementation but also autonomous reasoning about global structure, inter-function dependencies, and consistent state management without pre-supplied scaffolding. 

The injection-based evaluation paradigm enables SolEval to use execution-based metrics, such as Pass@k, because a complete compilation environment is guaranteed in advance. We argue that such metrics become less suitable as we move toward holistic repository-level code generation. Successful compilation requires LLMs not only to generate syntactically and semantically correct code, but also to correctly manage complex dependency structures. This challenge is illustrated by our benchmark dataset, which shows that the average code length exceeds 1,700 tokens and includes  cross-file dependencies, audited library imports (e.g., OpenZeppelin), shared interfaces, and compiler-version constraints. Under these conditions, even logically correct code may fail to compile due to missing or incompatible dependencies, a phenomenon we refer to as the \textit{compilation gap}. This gap fundamentally restricts the applicability of execution-based evaluation metrics, as compilation failure does not necessarily indicate semantic incorrectness. Consequently, evaluating repository-level Solidity code generation demands semantic-aware evaluation approaches that go beyond binary compilation success.

Overall, existing benchmarks and evaluation paradigms fall short of capturing the full complexity of repository-level Solidity smart contract generation. This gap motivates the need for new datasets, evaluation metrics, and empirical analyses specifically designed to assess how LLMs perform under realistic architectural and dependency constraints, which we aim to address in this work.

\subsection{Adaptation Strategies for LLM-driven Code Generation}

To bridge the gap between pre-trained LLMs and downstream tasks, particularly code generation, a range of model adaptation techniques has been widely explored. Rather than retraining models from scratch, these techniques aim to incorporate task-specific knowledge into LLMs and can be broadly categorized into prompting-based non-parametric approaches and parameter-updating approaches.

Prompting-based adaptation strategies modify the input context provided to the model at inference time, guiding generation through carefully designed instructions or examples. Zero-shot prompting relies solely on natural language task descriptions, while more structured techniques introduce explicit reasoning steps or auxiliary context. Chain-of-Thought (CoT) prompting encourages models to generate intermediate reasoning traces before producing final outputs, thereby improving performance on tasks that require multi-step logical reasoning~\cite{2.2:cot/Wei0SBIXCLZ22}. However, traditional free-text CoT often encounters a semantic gap, where the reasoning process diverges from the resulting code implementation when handling structured code data. To address this issue, Li \etal~\cite{DBLP:COT0} proposed Structured CoT (SCoT), which explicitly aligns reasoning steps with program control flows such as loops and branches, thereby improving logical rigor. Similarly, Li \etal~\cite{2.2:cot/LiQGCWTLWZ25} introduced the Chain of Functional Triggers (CoFT) strategy, which clarifies the functional semantics of key steps by incorporating standard programming identifiers. Collectively, these studies highlight that explicit structural guidance is critical for bridging natural language intent and formal code logic. This is also a vital requirement for smart contract generation, where error tolerance is low.

In-Context Learning (ICL) enables models to adapt to specific tasks by including relevant examples directly within the prompt~\cite{2.2:icl/BrownMRSKDNSSAA20}. Kapu \etal~\cite{2.2:icl/abs-2411-00865} proposed the DemoCraft framework, which improves generation accuracy by employing a specialized retrieval strategy to select high-quality examples. Meanwhile, Yang \etal~\cite{2.2:icl/YangCGLHLX25} found that clear variable and function naming is a key factor for effective examples, often more important than code formatting. Similarly, Patel \etal~\cite{2.2:icl/Patel24} evaluated the efficacy of ICL for library learning, demonstrating that providing API definitions in context enables models to use unseen or private libraries without fine-tuning. However, they also observed that ICL performance is sensitive to the inclusion of irrelevant APIs in the prompt. More generally, ICL is constrained by the model’s context window, which limits its ability to incorporate and leverage the extensive domain knowledge required for complex generation tasks.

Retrieval-Augmented Generation (RAG) complements prompting by dynamically retrieving relevant external knowledge, such as code examples or API documentation, and incorporating it into the input context, thereby reducing reliance on parametric memory alone. By augmenting the prompt with the retrieved information, RAG mitigates the fixed-context bottleneck inherent to standard prompting approaches~\cite{DBLP:RAG0}. Li \etal~\cite{2.2:rag/LiWWS25} proposed a framework based on multiple retrievers that improves generation quality by combining code features at different levels. From a context scope perspective, Gu \etal~\cite {2.3:Dsl/GuCLHZWWXW25} demonstrated the value of retrieving project-level APIs, highlighting that precise contextual information is essential to reduce hallucinations. Extending this idea to the repository level, Zhang \etal~\cite{2.2:rag/Zhang25} introduced a method to retrieve relevant knowledge distributed throughout an entire codebase, thus improving consistency in large projects. Despite the strong empirical performance of RAG, the risk that the retrieved code snippets may contain vulnerabilities, and thus introduce new security risks in security-sensitive domains, remains an unresolved challenge~\cite{2.2:rag/ThakurAPTDKG24}.

When general-purpose models underperform in specialized domains, supervised fine-tuning (SFT) remains a popular approach to integrate domain knowledge. Given the high computational cost of full-parameter fine-tuning, Parameter-Efficient Fine-Tuning (PEFT) techniques, such as Low-Rank Adaptation (LoRA)~\cite{DBLP:SFT0}, have emerged as practical alternatives. Weyssow \etal~\cite{2.2:rag/WeyssowZKLS25} demonstrated that updating only a small subset of parameters can achieve performance comparable to full-parameter fine-tuning for LLM-based code generation. Furthermore, Zhang \etal~\cite{2.2:ft/Zhang25} proposed an adaptive preference optimization approach, showing that targeted training guided by error logs can effectively mitigate specific categories of model errors.

\subsection{Domain-specific Code Generation}
Recent research has explored domain-specific code generation across a wide range of application areas. Gu \etal~\cite{2.3:Dsl/GuCLHZWWXW25} systematically evaluated LLM-based code generation in web and game development domains, showing that leveraging API knowledge and CoT can improve performance. Zhao \etal~\cite{2.3:Dsl/ZhaoLLWZ25} examined scientific computing programs and identified challenges related to specialized APIs, demonstrating that fine-tuning and few-shot learning can mitigate these issues. Extending to low-resource settings, Yang \etal~\cite{2.3:Dsl/YangKSZLJLL24} proposed MetaCoder, which applies meta-learning to improve code generation for domain-specific neural code tasks.

Domain-specific constraints are particularly pronounced in safety- and system-critical settings. In the aerospace domain, He \etal~\cite{2.3:Dsl/Rui2025} constructed a benchmark for embedded device code generation and found that general-purpose models exhibit high error rates when generating code compliant with the DO-178C safety standard, whereas domain-specific fine-tuning improves both functional correctness and compliance. Similarly, Vale \etal~\cite{2.3:Dsl/Pedro2023} studied industrial robotics and emphasized the necessity of modeling system-level factors such as timing constraints and hardware resource limits. In hardware description languages, Thakur \etal~\cite{2.2:rag/ThakurAPTDKG24} introduced VeriGen and demonstrated that domain-specific fine-tuning substantially improves functional correctness.

Another line of work focuses on the enforcement of structural and syntactic constraints in domain-specific code generation. Kang \etal~\cite{2.3:Dsl/KangCJJLCL25} showed that combining retrieval-augmented generation with preference optimization improves structural consistency in visual program generation. To directly address syntax constraints in DSLs, Wang \etal~\cite{2.3:Dsl/WangW0CSK23} proposed grammar prompting, which embeds Backus–Naur Form (BNF) specifications into prompts to guide LLMs toward syntactically valid outputs.

\section{Benchmark Construction and Evaluation Metrics}
\label{chap:chap3}
    \subsection{Overview}
\label{chap:3.1}
\begin{figure}[!htbp]
  \centering
  \includegraphics[width=\linewidth]{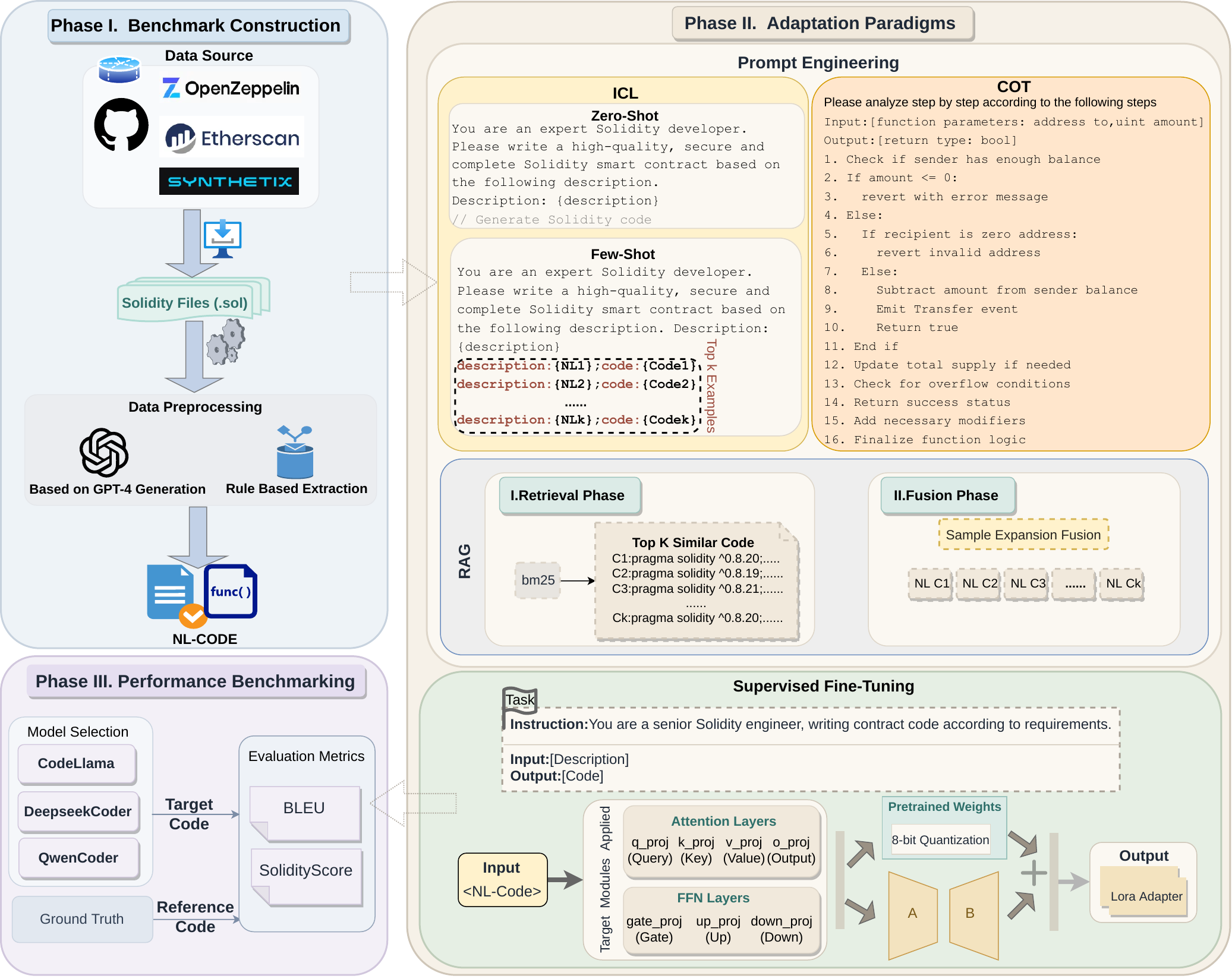}
  \caption{Overall framework of the empirical study on repository-level Solidity code generation}
  \label{fig:fig1-overview}
\end{figure}

As illustrated in Fig.~\ref{fig:fig1-overview}, this study employs a systematic three-phase framework to evaluate the effectiveness of LLMs in generating repository-level Solidity code. The process begins with Phase I (Benchmark Construction), during which we establish a high-quality dataset by collecting, cleaning, and synthesizing multi-source data into aligned natural language–code pairs. This is followed by Phase II (Adaptation Paradigms), which investigates diverse strategies ranging from inference-time prompt engineering (e.g., CoT, ICL, RAG) to parameter-efficient SFT. Finally, in Phase III (Performance Benchmarking), we utilize multiple adaptation paradigms across three representative baseline models to generate repository-level Solidity code. The quality of the generated code is then assessed through multi-dimensional evaluation metrics by comparing model outputs against ground-truth references.


\subsection{Benchmark Construction}
\label{chap:3.2}


Our benchmark is constructed from three Solidity code sources, namely Synthetix, OpenZeppelin, and Etherscan. They are selected according to the principle of triangulation \cite{Miller08} to ensure authority, diversity, and representativeness of real-world smart contract development. Synthetix is a large-scale DeFi protocol whose production-grade contracts implement complex financial logic and interdependent protocol mechanisms, making it suitable for evaluating a model’s ability to reason about sophisticated financial operations encoded in Solidity contracts~\cite{3.2:Synthetix/Werner22}. OpenZeppelin is an open-source smart contract security framework and tooling ecosystem for Ethereum and other EVM-compatible blockchains. It provides a widely adopted collection of reusable, modular, and extensively audited smart contract components, serving as a reference for security and standardization practices in the Ethereum ecosystem~\cite{2.1:KhanDVM23}. In contrast, Etherscan hosts a large repository of fully deployed, application-specific smart contracts sourced from the public Ethereum blockchain~\cite{3.2:Etherscan/Durieux20}. These contracts span a wide range of domains, coding styles, and quality levels, reflecting the heterogeneity and practical constraints of real-world deployments rather than curated best practices. We curated the Synthetix and OpenZeppelin corpora from their official version-controlled repositories. Etherscan contracts were collected from their verified contract registry. A summary of each source, including popularity metrics, is provided in Table~\ref{tab:dataset}.



\begin{table}[!h]
  \caption{Statistics of target platforms and data samples in SolidityBench}
  \label{tab:dataset}
  \centering
  \resizebox{\linewidth}{!}{
  \begin{tabular}{llccr} 
    \toprule
    \textbf{Platform} & \textbf{Domain} & \textbf{Description} & \textbf{Stars} & \textbf{Samples} \\
    \midrule
    Synthetix & DeFi/Finance & Complex decentralized synthetic asset protocol & 1.3K & 550 \\
    OpenZeppelin & Security & Audited, standardized smart contract libraries & 26.6K & 1,150 \\ 
    Etherscan & Diverse Apps & Large-scale verified real-world deployed contracts & N/A & 3,770 \\
    \midrule
    \textbf{Total} & & & & \textbf{5,470} \\ 
    \bottomrule
  \end{tabular}
  }
\end{table} 

To process the raw code, we developed a rigorous automated cleaning pipeline. First, we performed fine-grained filtering and removal of code comments using regular expressions. The operation removed non-functional boilerplate text, such as copyright notices and licenses, while precisely preserving NatSpec tags (e.g., @dev and @param) which encapsulate high-level semantic specifications and serve as critical semantic anchors for subsequent natural language construction. Next,  to mitigate stylistic inconsistencies arising from diverse programming conventions, all code samples were standardized by reformatting them in strict accordance with the official Solidity style guide. This stem unified indentation, line breaks, and code layout specifications, ensuring the model could focus on learning the code's deep semantic structures.


\begin{figure}[!htbp]
    \centering
    \begin{tcolorbox}[
        colback=gray!5!white,
        colframe=gray!75!black,
        width=0.8\linewidth,  
        title=\textbf{GPT-4 Prompt Template for Solidity Description Generation}
    ]
    \small
    \textbf{\# System Instruction} \\
    You are an expert Solidity smart contract auditor and developer. Your task is to generate a high-quality, repository-level functional description for the provided Solidity code snippet.

    \smallskip

    \textbf{\# Constraints \& Style Guidelines}
    \begin{itemize}[leftmargin=*, noitemsep, topsep=2pt, partopsep=0pt]
        \item \textbf{Format}: Strictly follow \texttt{NatSpec} style (\texttt{@dev}, \texttt{@notice}).
        \item \textbf{Content}: Focus on logic intent, state updates, and control flow.
        \item \textbf{Security}: Explicitly mention patterns (e.g., \texttt{nonReentrant}, \texttt{onlyOwner}).
        \item \textbf{Tone}: Maintain a professional technical tone (OpenZeppelin style).
    \end{itemize}

    \smallskip

    \textbf{\# Few-Shot Demonstration} \\
    \textit{<Input Code>}: \\
    \texttt{function withdraw(uint256 amount) public nonReentrant \{ ... \}} \\
    \textit{<Target Output>}: \\
    {\raggedright 
    \texttt{@notice Allows users to withdraw funds. @dev Follows Checks-Effects-Interactions pattern to prevent re-entrancy during external call.}
    \par} 

    \smallskip

    \textbf{\# Current Task} \\
    \textit{Input:} \texttt{\{Input\_Solidity\_Code\}} \\
    \textit{Output:} 
    \end{tcolorbox}
    \vspace{-0.2cm}
    \caption{The GPT-4 prompt template designed for generating natural language descriptions of Solidity code.}
    \label{fig:GPT4-prompt}
\end{figure}
The alignment of high-quality natural language descriptions with code is crucial to code generation and a decisive factor in the effectiveness of SFT. However, raw smart contract code is often plagued by a paucity of high-quality natural language documentation, with descriptions that are either missing or ambiguous. To address this, we employed a hybrid labeling strategy that fuses rule-driven extraction with LLM-based semantic completion ~\cite{3.2:Magicoder/Wei20,3.2:Self-Instruct/Wang23}.  Specifically, for samples adhering to Solidity’s Natural Language Specification Format (NatSpec), accounting for  approximately 65\%,  we applied regular expressions to filter out irrelevant text, such as licenses, and extracted key semantic tags (e.g., @dev and @param) to construct high-quality, repository-level Solidity code descriptions.  For the remaining 35\%  lacking high-quality descriptions, GPT-4 was employed to generate missing code descriptions. To ensure high consistency in style and semantics between the generated content and native NatSpec annotations, we designed instruction templates based on few-shot learning and performed iterative optimization against existing high-quality samples. This guided  GPT-4 to synthesize functional descriptions that strictly adhere to semantic accuracy and stylistic norms. The specific prompt template, incorporating these constraints and few-shot examples, is presented in Fig.~\ref{fig:GPT4-prompt}. 

Finally,  we constructed a high-quality domain-specific dataset containing 5,470 smart contracts. In accordance with standard machine learning evaluation  protocols~\cite{3.2:CodeSearchNet/Husain19,3.2:CodeBERT/Feng20,3.2:CodeXGLUE/Lu21}, the dataset was strictly partitioned into training, validation, and test sets in an 8:1:1, comprising 4,376, 547, and 547 samples respectively. This partitioning strategy ensures sufficient model fitting while providing a balanced basis for hyperparameter optimization and unbiased evaluation. 

Statistical results are summarized in Table~\ref{tab:dataset_stats}, which shows that the average code length exceeds 1,737 tokens, and the average number of functions per contract amounts to 19.09. Such high functional density demonstrates that our dataset encompasses diverse and complex real-world business logic.  
To further elaborate on the characteristics of the dataset, an example from SolidityBench is shown in Fig.~\ref{fig:data_sample}. Distinct from trivial function completion tasks~\cite{intro:soleval/Peng25}, This example concretely reveals three inherent and critical challenges posed by our task: (1) \textbf{Dependency Management}, where the model is required to correctly handle external imports (e.g., OpenZeppelin libraries) for the reuse of audited and verified code;
(2) \textbf{Architectural Integrity}, which demands the correct implementation of inheritance hierarchies (e.g., inheriting \texttt{Ownable} and \texttt{ReentrancyGuard});
(3) \textbf{Global Coordination}, where the model must generate multiple interrelated functions (e.g., \texttt{stake}, \texttt{withdraw}, and \texttt{claimReward}) that consistently operate on shared global states. 
This observation demonstrates that our benchmark entails rigorous demands on the model’s capacity to understand file-level contextual information and global logical flow.
\begin{table}[H]
\centering
\caption{Dataset partitioning and key statistical characteristics. The ``Avg. Func.'' column represents the average number of functions per contract}
\label{tab:dataset_stats}
\renewcommand{\arraystretch}{1} 
\setlength{\tabcolsep}{3pt}       

\begin{tabular}{l c c c c c}
\toprule
\textbf{Dataset} & \textbf{Samples} & \textbf{Proportion} & \textbf{\makecell{Avg. Code Length \\ (tokens)}} & \textbf{\makecell{Avg. Desc. Length \\ (tokens)}} & \textbf{\makecell{Avg. Func. \\ (per file)}} \\
\midrule
Training Set   & 4,376 & 80\%  & 1,746.45 & 61.75 & 19.45 \\
Validation Set & 547   & 10\%  & 1,789.72 & 61.40 & 18.62 \\
Test Set       & 547   & 10\%  & 1,614.16 & 61.40 & 16.63 \\
\midrule
\textbf{Total} & \textbf{5,470} & \textbf{100\%} & \textbf{1,737.55} & \textbf{61.68} & \textbf{19.09} \\
\bottomrule
\end{tabular}
\end{table} 
\begin{figure}[!htbp]
    \centering
    \begin{tcolorbox}[
        colback=gray!5!white,      
        colframe=gray!75!black,    
        width=\linewidth,          
        arc=2mm,                   
        boxrule=1pt,               
        title=\textbf{An Examplar Sample from SolidityBench} 
    ]
        
        \small
        \textbf{[Natural Language Description]} \\
        \textit{A staking contract that allows users to stake tokens and earn rewards. It imports the OpenZeppelin ERC20 library for token handling and Ownable for access control. The contract includes functions for staking, withdrawing, and claiming rewards, ensuring protection against re-entrancy.}
        
        \smallskip 
        
        \textbf{[Solidity Code (Ground Truth)]}
        
        \begin{lstlisting}[
            language=Solidity, 
            basicstyle=\scriptsize\ttfamily, 
            backgroundcolor=\color{gray!5!white}, % 确保代码背景与盒子背景一致
            frame=none, 
            aboveskip=2pt, 
            belowskip=0pt
        ]
pragma solidity ^0.8.0;
import "@openzeppelin/contracts/token/ERC20/IERC20.sol";
import "@openzeppelin/contracts/access/Ownable.sol";
import "@openzeppelin/contracts/security/ReentrancyGuard.sol";
contract StakingContract is Ownable, ReentrancyGuard {
    IERC20 public stakingToken;
    mapping(address => uint256) public balances;
    // ... (State Variables) ...
    function stake(uint256 amount) external nonReentrant {  // ... (Logic for staking) ...      }
    function withdraw(uint256 amount) external nonReentrant {  // ... (Logic for withdrawing) ...     }
    function claimReward() external { // ... (Logic for rewards) ... }
}
        \end{lstlisting}
    \end{tcolorbox}
    \vspace{-0.2cm}
    \caption{An example from SolidityBench, which illustrates a natural language description aligned with a full smart contract involving external imports and inheritance hierarchies.}
    \label{fig:data_sample}
\end{figure}


\subsection{Evaluation Metrics}
\label{chap:3.3}
While execution-based evaluation is valuable for assessing LLM-driven code generation, it is not always feasible as a primary metric for repository-level Solidity code generation. This is due to environment dependencies, deployment context, and the high prevalence of partial or uncompilable outputs, which can prevent reliable execution even when the generated code is logically meaningful. Given the availability of human-written ground-truth smart contracts, we therefore adopt text-based evaluation metrics, measuring the similarity between LLM-generated code and reference implementations to assess the semantic correctness of generated outputs. In addition, we treat compilation success as an auxiliary signal for analyzing the challenges of Solidity code generation (ref. Section~\ref{sec:rq4}), complementing the primary text-based semantic evaluation.

We retain the widely used BLEU metric as a lightweight indicator of surface-form consistency. Solidity is a strongly typed language with strict syntactic requirements, including pragma and version declarations, function signatures, and precise use of punctuation and delimiters. In this setting, $n$-gram overlap serves as a practical proxy for whether a model reproduces essential syntactic scaffolding and conventional code patterns observed in human-written contracts. However, BLEU is inherently limited in its expressiveness: it rewards token-level co-occurrence rather than semantic equivalence and penalizes logically correct variants that differ due to variable renaming, control-flow reorganization, or alternative but functionally equivalent implementations. As a result, BLEU alone is insufficient for assessing the functional correctness of generated Solidity code.

To capture semantic correctness beyond lexical overlap, we examined existing semantic code evaluation metrics, including CodeBLEU~\cite{DBLP:journals/corr/CodeBleu} and CodeBERTScore~\cite{DBLP:conf/emnlp/CodeBERTScore}, but found them insufficient for Solidity in practice. CodeBLEU relies heavily on parsing-driven signals such as ASTs, which makes it brittle in the Solidity setting: even minor syntactic errors can prevent successful parsing, resulting in missing or invalid scores precisely for partially correct outputs where robust evaluation is most needed. CodeBERTScore, while more tolerant of imperfect syntax, depends on encoders pre-trained primarily on general-purpose programming languages. This mismatch introduces a domain gap that limits its ability to accurately represent Solidity-specific semantics, particularly security-critical constructs (e.g., access control modifiers and reentrancy patterns) and execution constraints such as gas-related considerations.

To overcome these limitations, we propose SolidityScore, a domain-adapted semantic evaluation metric that builds upon the core matching principle of CodeBERTScore while tailoring both representation and weighting to Solidity. Specifically, the metric encodes candidate and reference code into contextual token embeddings, computes pairwise cosine similarities between tokens, and aggregates alignment using a greedy matching strategy to produce an F1-style similarity score. We introduce two key adaptations: (1) replacing the general-purpose encoder with Solidity-LLM\footnote{\url{https://huggingface.co/Chain-GPT/Solidity-LLM}}, a model fine-tuned on 650K high-quality Solidity instruction pairs, to capture domain-specific semantics; and (2) substituting standard inverse document frequency (IDF) weighting with a domain-adaptive weighting scheme that prioritizes security- and logic-critical tokens based on their functional importance rather than statistical rarity. 

Specifically, \textit{SolidityScore} is computed via the following three steps:

\textbf{Token Representation.} To capture the functional semantics, we encode the natural language description along with the contract code. Formally, let $x = \langle x_1, \dots, x_k \rangle$ represent the sequence of the natural language requirement, $y^* = \langle y^*_1, \dots, y^*_m \rangle$ denote the ground-truth reference contract, and $\hat{y} = \langle \hat{y}_1, \dots, \hat{y}_n \rangle$ denote the generated candidate contract. The description $x$  is concatenated separately with the reference code $y^*$  and the candidate code $\hat{y}$. The resulting sequences are then tokenized using the tokenizer $\mathcal{T}_{\mathcal{M}}$ of Solidity-LLM model to produce the  corresponding input sequences:
\begin{equation}
\begin{aligned}
\mathcal{T}_{\mathcal{M}}(x \cdot y^*) &= \langle x_1, \dots, x_k, y^*_1, \dots, y^*_m \rangle \\
\mathcal{T}_{\mathcal{M}}(x \cdot \hat{y}) &= \langle x_1, \dots, x_k, \hat{y}_1, \dots, \hat{y}_n \rangle
\end{aligned}
\label{eq:tokenization}
\end{equation}
where  $k$, $n$, and $m$ represent the number of tokens in the natural language description, the reference contract, and the generated candidate contract, respectively. These sequences are fed into Solidity-LLM, from which we extract context-aware semantic vectors from its final hidden layer. We denote the vector of the $i$-th token in the reference code as $\mathbf{v}_{y^*_i}$ and  that of the $j$-th token in the candidate code as $\mathbf{v}_{\hat{y}_j}$.

\textbf{Similarity Computation.} We compute the cosine similarity (defined in Eq.~\ref{eq:sim}) between token embeddings to measure semantic relatedness at the token level. This similarity reflects local semantic alignment between individual tokens and does not yet incorporate sequence-level structure. 
\begin{equation}
\text{sim}(\mathbf{u}, \mathbf{v}) = \frac{\mathbf{u}^\top \mathbf{v}}{\|\mathbf{u}\| \|\mathbf{v}\|}
\label{eq:sim}
\end{equation}

\textbf{SolidityScore Calculation.} Domain-weighted Recall ($R_{sol}$) and Precision ($P_{sol}$) are obtained by greedy matching over the token-wise similarity matrix. During this matching process, we apply a domain-adaptive weight function $w(\cdot)$ to individual tokens to reflect their functional importance. Specifically, the weights are derived from IDF statistics computed over the Solidity training corpus constructed in this study, assigning higher importance to tokens that are critical in Solidity. The final SolidityScore is defined as their harmonic mean (F1-score). Recall (Eq. (\ref{eq:recall})) measures how well the reference logic is covered by the generated code.  Precision (Eq. (\ref{eq:precision})) assesses the relevance of the generated tokens with respect to the reference, and the combined SolidityScore (Eq. (\ref{eq:f1})) provides a balanced, length-invariant evaluation.
\begin{equation}
      R_{sol} = \frac{1}{\sum_{i=1}^{m} w(y^*_i)} \sum_{i=1}^{m} w(y^*_i) \cdot \max_{j} \text{sim}(\mathbf{v}_{y^*_i}, \mathbf{v}_{\hat{y}_j}) \label{eq:recall} \\[10pt]
\end{equation}
  
\begin{equation}   
   P_{sol} = \frac{1}{\sum_{j=1}^{n} w(\hat{y}_j)} \sum_{j=1}^{n} w(\hat{y}_j) \cdot \max_{i} \text{sim}(\mathbf{v}_{\hat{y}_j}, \mathbf{v}_{y^*_i}) \label{eq:precision} \\[10pt]
\end{equation}

\begin{equation}   
     \text{SolidityScore} = 2 \cdot \frac{P_{sol} \cdot R_{sol}}{P_{sol} + R_{sol}} \label{eq:f1}
\end{equation}

SolidityScore produces values in the range [0,1]. A high SolidityScore indicates that the generated code preserves critical business logic and security-relevant constraints even when syntactic variations (e.g., variable renaming or code reorganization), which would typically lead to low scores under surface-form metrics like BLEU.

In summary, we pair BLEU with SolidityScore to obtain a balanced evaluation of Solidity code generation. BLEU captures syntactic and stylistic proximity to the reference implementation, while SolidityScore measures semantic alignment that is robust to surface-level differences. High scores on both metrics suggest code that is both structurally conventional and logically faithful. Conversely, low BLEU but high SolidityScore often reflects functionally correct solutions expressed through alternative naming conventions or reorganized code structures.

\section{Experimental Setup}
\label{chap:chap4}
To ensure the reliability and reproducibility of the experimental results, all experiments were conducted on a unified hardware and software platform. A high-performance computing node  equipped with a single NVIDIA GeForce RTX 5090 GPU (32GB VRAM) provided the computational resources for model inference and training. The software environment was built upon Python 3.12 and the PyTorch 2.8.0 framework. 
\subsection{Model Selection}
\label{chap:4.1}

To ensure a comprehensive and representative evaluation, we selected three popular open-source LLMs recognized for their strong coding performance: CodeLlama~\cite{DBLP:journals/corr/CodeLlama}, DeepSeek-Coder~\cite{DBLP:journals/DeepSeek-Coder}, and Qwen2.5-Coder~\cite{DBLP:journals/Qwen2.5-Coder}. Specifically, we use CodeLlama-7B-Instruct-HF, DeepSeek-Coder-6.7B-Instruct, and Qwen2.5-Coder-7B-Instruct, all downloaded from their official standard releases on the Hugging Face Hub.

The selection of LLMs was guided by four primary criteria:

(1) Representative paradigms: The chosen models reflected the predominant training paradigms in code-specialized LLMs. CodeLlama exemplifies domain adaptation of a general-purpose foundation model. DeepSeek-Coder represents a model pre-trained from scratch on code, while Qwen2.5-Coder illustrates a powerful generalist model with robust coding capabilities. 

(2) Controlled model capacity: We standardized our comparison at the 7-billion (7B) parameter  scale. While ICL permits the use of significantly larger models, we restrict our baselines to 7B to ensure a fair, apples-to-apples comparison of learning paradigms (SFT vs. ICL) under identical model capacities. Furthermore, the 7B scale represents the practical sweet spot for local deployment on consumer-grade hardware \cite{Zhao0CWAT24}.

(3) Accessibility and reproducibility: All selected models were fully open-source and publicly available on the Hugging Face Hub\footnote{\url{https://huggingface.co/}}, ensuring experimental transparency and facilitating community replication. 

(4) Established baselines: These models consistently rank at the top of major code generation benchmarks (e.g., HumanEval~\cite{DBLP:journals/CODEX} and MBPP~\cite{DBLP:journals/MBPP}) and are frequently adopted as reference models in contemporary~\cite{4.1:NaturalCodeBench/Zhang24,4.1:pro/Yu25,4.1:journals/corr/Rahman25}, solidifying their status as standard benchmarks for comparison.

\subsection{Research Question}
\label{chap:4.2}

Using the evaluation framework introduced in Section~\ref{chap:chap3}, we assess existing LLMs on repository-level Solidity code generation through the following five research questions (RQs).

\begin{itemize}
\item \textbf{RQ1: } How do general-purpose LLMs perform in generating repository-level Solidity code under a zero-shot setting?  

\item \textbf{RQ2: } To what extent can prompting-based adaptation strategies improve repository-level Solidity code generation? 
\begin{itemize}
\item \textbf{RQ2.1:} Can structured reasoning via CoT improve repository-level Solidity code generation capabilities?
            
\item \textbf{RQ2.2:} Can contextual demonstrations via ICL improve repository-level Solidity code generation capabilities?
            
\item \textbf{RQ2.3:} Can dynamic knowledge injection via RAG improve repository-level Solidity code generation capabilities?
\end{itemize}

\item \textbf{RQ3: } How does domain-specific SFT compare with prompt-based strategies for repository-level Solidity code generation? 

\item \textbf{RQ4: } Does \textit{SolidityScore} provide a more reliable assessment of the semantic correctness of generated Solidity code than BLEU?

\item  \textbf{RQ5:} To what extent can LLMs generate compilable repository-level Solidity contracts, and what are the dominate types of compilation errors? 
\end{itemize}

These RQs are designed to form a coherent progression: from establishing baseline capabilities, to examining different adaptation paradigms, to validating evaluation metrics and practical feasibility. Specifically, RQ1 establishes the zero-shot performance of general-purpose LLMs without any form of adaptation. RQ2 investigates inference-time adaptation via prompt engineering strategies, with sub-questions (RQ2.1–RQ2.3) isolating the effects of individual mechanisms. RQ3 then examines parameter-level adaptation by comparing domain-specific supervised fine-tuning against prompt-based approaches. Beyond performance comparison, RQ4 focuses on evaluation validity by assessing whether our proposed semantics-aware metric, SolidityScore, provides more reliable semantic assessment than BLEU. Finally, RQ5 complements the semantic evaluation by examining the compilation feasibility of generated code and diagnosing the types of compilation failures that arise.

\subsection{Hyperparameter Setting }
\label{chap:4.3}

The effectiveness of zero-shot and various prompt engineering strategies was  empirically evaluated. To ensure the stability and reproducibility of the generated results, a low-temperature sampling strategy (with the temperature coefficient $T$ uniformly set to 0.2) was adopted across all inference experiments. This setting significantly reduced stochastic fluctuations while preserving a minimal degree of diversity, thereby providing a fair benchmark for performance comparison across different strategies.

For the supervised fine-tuning (SFT) experiments, we employed LoRA for parameter-efficient fine-tuning, considering computational constraints and training efficiency. Specifically, the training duration was set to 30 epochs using a cosine learning rate scheduler, with an initial learning rate of 3e-4 and a warmup ratio of 0.05~\cite{4.3:SGDR/Loshchilov17}. To balance training stability and resource usage, the fine-tuning process utilized a batch size of 2 with a gradient accumulation of 5. Furthermore, an early stopping mechanism with a patience value of 2 was implemented to mitigate overfitting risks, following established regularization strategies~\cite{4.3:EarlyStopping/Dodge20,4.3:EarlyStopping/Mosbach21,4.3:EarlyStopping/Prechelt96}. This configuration yielded stable convergence and consistent performance in our preliminary tuning experiments.

\section{Experimental Results and Analysis}
\label{chap:chap5}
\subsection{RQ1: Evaluating LLMs for Solidity Code Generation under Zero-Shot Setting}

\leavevmode\noindent \textbf{Methodology.}
To systematically evaluate the performance of general-purpose LLMs in generating repository-level Solidity code, we conduct experiments under a zero-shot setting. Under this setup, models are provided only with natural language functional descriptions from the test set, without any code demonstrations, reasoning cues, or auxiliary context. We designed a unified instruction template for all models, explicitly requiring the generation of complete, compilable, and syntactically correct smart contracts based on the given descriptions. The designed template is shown in Fig. \ref{fig:zeroshot_prompt}. The performance of the studied LLMs is quantitatively and qualitatively analyzed under this scenario. Quantitatively, we utilize the BLEU metric to measure lexical overlap and syntactic adherence, and employ SolidityScore to assess semantic alignment and logical consistency within the Solidity domain; specifically, the mean values of both metrics across all samples in the test set are reported.  Qualitatively, a manual inspection is conducted based on 5 instances randomly sampled from the test set. We specifically check compliance with Solidity-specific constraints, such as the correct use of modifiers and gas cost optimization, as well as critical security patterns including the CEI pattern. 

\begin{figure}[h]
\centering
\begin{tcolorbox}[
    colback=gray!5!white,
    colframe=gray!75!black,
    width=\linewidth,
    title=\textbf{Zero-shot Prompt Template}
]
\small
\textbf{\# Instruction} \\
You are an expert Solidity developer. Please write a high-quality, secure, and complete Solidity smart contract based on the following description.

\smallskip

\textbf{\# Task Specification} \\
\textit{Description:} \texttt{\{description\}}

\smallskip

\textbf{\# Response} \\
\textit{Code:}
\end{tcolorbox}
\vspace{-0.2cm}
\caption{The zero-shot prompt template used for Solidity code generation.}
\label{fig:zeroshot_prompt}
\end{figure}

\noindent\textbf{Results.}
The quantitative results in Table \ref{tab:zero_shot_results} show clear performance differences among the evaluated models under the zero-shot setting. Qwen2.5-Coder-7B-Instruct achieved the highest scores, with a BLEU of 18.84 and a SolidityScore of 0.5566, demonstrating strong foundational generation capabilities. DeepSeek-Coder-6.7B-Instruct delivered moderate performance, with scores of 12.14 and 0.5311, respectively. In contrast, CodeLlama-7B-Instruct-HF showed significantly lower proficiency, obtaining only 7.99 in BLEU and 0.4692 in SolidityScore, showing limited ability to directly generate valid repository-level Solidity  code under the scenario. 

\begin{table}[H]
\centering
\caption{Performance evaluation of models in the zero-shot setting}
\label{tab:zero_shot_results}
\renewcommand{\arraystretch}{1.2}
\begin{tabular}{ccc}
\toprule
\textbf{Model} & \textbf{BLEU} & \textbf{SolidityScore} \\
\midrule
CodeLlama-7B-Instruct-HF      & 7.9957  & 0.4692 \\
Deepseek-Coder-6.7B-Instruct  & 12.1412 & 0.5311 \\
Qwen2.5-Coder-7B-Instruct     & \textbf{18.8433} & \textbf{0.5566} \\
\bottomrule
\end{tabular}
\end{table} 

In addition to the quantitative evaluation, we performed a manual inspection by randomly sampling 5 instances from the test set to assess the functional and security quality of the generated code. This qualitative analysis showed that, under the zero-shot setting, all three models exhibited recurring and critical shortcomings. Common issues across models included the incorrect application of essential modifiers such as \texttt{payable}, the omission of visibility specifiers for state variables, and violations of foundational security patterns, for example, performing external calls before updating internal state. Although Qwen2.5-Coder achieved the highest quantitative scores, its generated code still contained significant deficiencies. Our inspection frequently identified syntactic violations and logical flaws in its outputs, primarily due to non-adherence to Solidity-specific domain rules and security conventions. These observations suggest that the unsatisfactory performance under the zero-shot setting may be attributable to missing domain-specific guidance, a hypothesis that we further examine through prompt-based and fine-tuning strategies in subsequent RQs.


\noindent\textbf{Answer to RQ1.}
General-purpose LLMs struggle to generate accurate repository-level Solidity code under a zero-shot setting, as reflected by consistently low BLEU and SolidityScore values. Although Qwen2.5-Coder achieves higher scores than DeepSeek-Coder and CodeLlama, all evaluated models exhibit pervasive syntactic and security-related defects, indicating that improvements in baseline model capability alone are insufficient to achieve reliable repository-level Solidity code generation.

\subsection{RQ2: Evaluating Different Prompt Optimization Strategies on Solidity Generation}
\subsubsection{RQ2.1: Effectiveness of Structured Chain-of-Thought} \mbox{} \\
\leavevmode\noindent \textbf{Methodology.}
The effectiveness of CoT prompting in code generation tasks has been extensively validated in existing literature. However, traditional free-form CoT is often limited by its unstructured reasoning process and insufficient alignment with the inherent structures of code. To address these challenges, Li \etal \cite{DBLP:COT0}  proposed a SCoT method, which explicitly guides reasoning through foundational programming constructs (e.g., sequences, branches, and loops) to enhance generation quality.

As a domain-specific languages for smart contract development, Solidity enforces strict architectural and security constraints. Generating correct code requires a thorough prior understanding of state variable relationships and access control logic. The SCoT approach mitigates common logical leaps and structural omissions by enforcing a stepwise reasoning process, progressing from requirement comprehension to structural planning and finally to implementation. This method mandates the explicit definition of control flows and input-output relationships, aligning closely with core Solidity constructs such as function signatures and state transitions. In this sense, following a previous study \cite{DBLP:COT0}, we designed a SCoT prompt template, as shown in Fig. \ref{fig:scot_prompt}, specifically for Solidity code generation. The template guides the model to first parse the requirements by identifying key business entities, then plan the structure of function interfaces and control flows, and finally generate concrete Solidity code based on this structured blueprint.
\begin{figure}[h]
\centering
\begin{tcolorbox}[
    colback=gray!5!white,
    colframe=gray!75!black,
    width=\linewidth,
    title=\textbf{SCoT Prompt Template}
]
\small
\textbf{\# Instruction} \\
You are a Solidity expert. Analyze the requirements step-by-step using Structured Chain-of-Thought before generating code.

\smallskip

\textbf{\# Reasoning Framework}
\begin{itemize}[leftmargin=*, noitemsep, topsep=2pt]
    \item \textbf{1. Interface Analysis}: Define inputs (e.g., \texttt{uint amount}) and outputs.
    \item \textbf{2. Control Flow Planning}:
    \begin{itemize}[leftmargin=1.5em, nosep]
        \item \textit{Sequential}: Check balance $\rightarrow$ Update state $\rightarrow$ Emit events.
        \item \textit{Branch}: Define conditions (e.g., \texttt{if amount <= 0}).
        \item \textit{Loop}: Define iteration logic or explicitly state "None".
    \end{itemize}
    \item \textbf{3. Security \& Finalization}: Verify against vulnerabilities (e.g., Re-entrancy) and apply modifiers (e.g., \texttt{onlyOwner}).
\end{itemize}

\smallskip

\textbf{\# Task} \\
\textit{Requirement:} \texttt{\{description\}} \\
\textit{SCoT Reasoning \& Code:}
\end{tcolorbox}
\vspace{-0.2cm}
\caption{The Structured Chain-of-Thought prompt template designed for Solidity code generation.}
\label{fig:scot_prompt}
\end{figure}




\noindent\textbf{Results.}
The experimental results in Table \ref{tab:scot_results} show that implementing the SCoT method substantially improved the  generation performance across all studied models. This outcome confirms the general effectiveness of explicit reasoning guidance in complex programming tasks. Regarding quantitative metrics, CodeLlama-7B-Instruct-HF showed the greatest improvement, with its BLEU score increasing from 7.99 to 17.59 and its SolidityScore rising from 0.4692 to 0.5502. DeepSeek-Coder and Qwen2.5-Coder also exhibited consistent performance uplifts. Qwen2.5-Coder maintained its leading position, achieving a BLEU score of 20.33 and a SolidityScore of 0.5619.
\begin{table}[H]
\centering
\caption{Performance evaluation of models in the SCOT setting}
\label{tab:scot_results}
\renewcommand{\arraystretch}{1.2}
\begin{tabular}{lcc}
\toprule
\textbf{Model} & \textbf{BLEU} & \textbf{SolidityScore} \\
\midrule
CodeLlama-7B-Instruct-HF      & 17.5946 & 0.5502 \\
Deepseek-Coder-6.7B-Instruct  & 14.7973 & 0.5309 \\
Qwen2.5-Coder-7B-Instruct     & \textbf{20.3323} & \textbf{0.5619} \\
\bottomrule
\end{tabular}
\end{table} 

Similar to RQ1, a manual inspection was conducted to qualitatively assess the quality of the generated code using the SCoT prompt, based on 5 instances randomly sampled from the test set. We found that the SCoT strategy effectively mitigated critical structural deficiencies identified in RQ1, such as missing constructor initializations and incorrect inheritance hierarchies, demonstrating a marked reduction in these architectural errors compared to the zero-shot baseline. With the introduction of structured planning steps, models demonstrated significantly improved accuracy in defining function visibility (e.g., external vs. public) and security-critical access control modifiers (e.g., onlyOwner) and a stronger adherence to standard conventions for declaring constructors and events. This qualitative analysis suggests that the reasoning chain not only refines local syntax, but also enhances the ’ high-level understanding of the overall contract architecture of the models. This architectural coherence is directly reflected in the increase in SolidityScore, which places a heavy weight on the correctness of state variable relationships and function interfaces. By mandating a deliberate design phase before code writing, SCoT ensures tight alignment between the overarching logic and its detailed implementation within Solidity’s constrained development paradigm.

\noindent\textbf{Answer to RQ2.1.}
SCoT outperforms the zero-shot baseline and effectively enables the mapping of natural language functional descriptions to Solidity’s rigorous syntactic and state-dependency constraints, thereby improving code generation quality. For models with lower baseline performance (e.g., CodeLlama), the structured reasoning steps primarily mitigated fundamental structural deficiencies, leading to observed improvements in syntactic coherence and completeness. In contrast, for models exhibiting stronger baseline capabilities (e.g., Qwen2.5), the explicit planning phase facilitated the accurate implementation of complex business logic and security patterns, such as adherence to the CEI pattern. Overall, decomposing the generation process reduces both low-level syntactic errors and high-level logical oversights, ensuring higher architectural integrity.

\subsubsection{RQ2.2: Effectiveness of in-context learning} 
\begin{figure}[h]
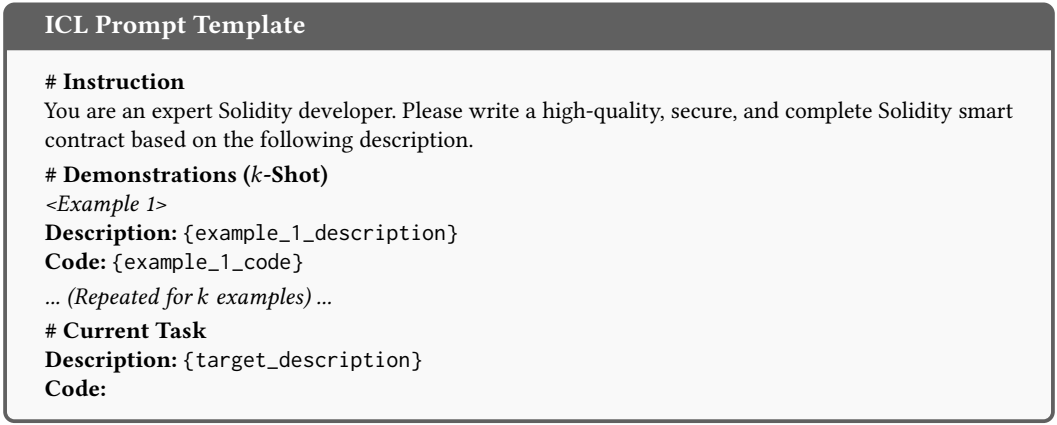

\centering
\begin{tcolorbox}[
    colback=gray!5!white,
    colframe=gray!75!black,
    width=\linewidth,
    title=\textbf{ICL Prompt Template}
]
\small
\textbf{\# Instruction} \\
You are an expert Solidity developer. Please write a high-quality, secure, and complete Solidity smart contract based on the following description.

\smallskip

\textbf{\# Demonstrations ($k$-Shot)} \\
\textit{<Example 1>} \\
\textbf{Description:} \texttt{\{example\_1\_description\}} \\
\textbf{Code:} \texttt{\{example\_1\_code\}}

\vspace{2pt}
\textit{... (Repeated for $k$ examples) ...}

\smallskip

\textbf{\# Current Task} \\
\textbf{Description:} \texttt{\{target\_description\}} \\
\textbf{Code:}
\end{tcolorbox}
\vspace{-0.2cm}
\caption{The In Context Learning prompt template designed for Solidity code generation.}
\label{fig:icl_prompt}
\end{figure}

\mbox{} \\
\leavevmode\noindent \textbf{Methodology.}
To explore the effectiveness and limitations of ICL for repository-level Solidity code generation, we randomly sampled contracts from the training set as few-shot demonstrations. These demonstrations were used to construct ICL prompt templates containing \(k\) examples, where \(k \in \{1, 2, 3, 4\}\). Random sampling was adopted to focus on the model’s intrinsic few-shot learning capability, without introducing confounding factors from external example selection mechanisms. In particular, by avoiding similarity-based retrieval or heuristic ranking strategies, this design ensures that observed performance differences reflect the model’s ability to learn from contextual examples rather than the effectiveness of a specific selection algorithm. The ICL prompt template we designed is shown in Fig. \ref{fig:icl_prompt}. 

This setup allows us to investigate two aspects of ICL behavior. First, we examine whether the inclusion of few-shot demonstrations improves the structural organization and stylistic consistency of generated Solidity code. Second, we analyze how model performance evolves as the number of demonstrations increases, assessing whether additional examples lead to continuous performance gains or instead result in diminishing returns or performance degradation. To control experimental variables, all models share the same pool of demonstration examples, with only the number of included examples varying across settings. This design allows performance changes to be primarily attributed to the model’s capacity to learn and generalize patterns from the provided context.

\begin{figure*}[h]
  \centering
  
  \subfigure[BLEU]{
    \label{fig:icl-bleu}
    \includegraphics[width=0.47\linewidth]{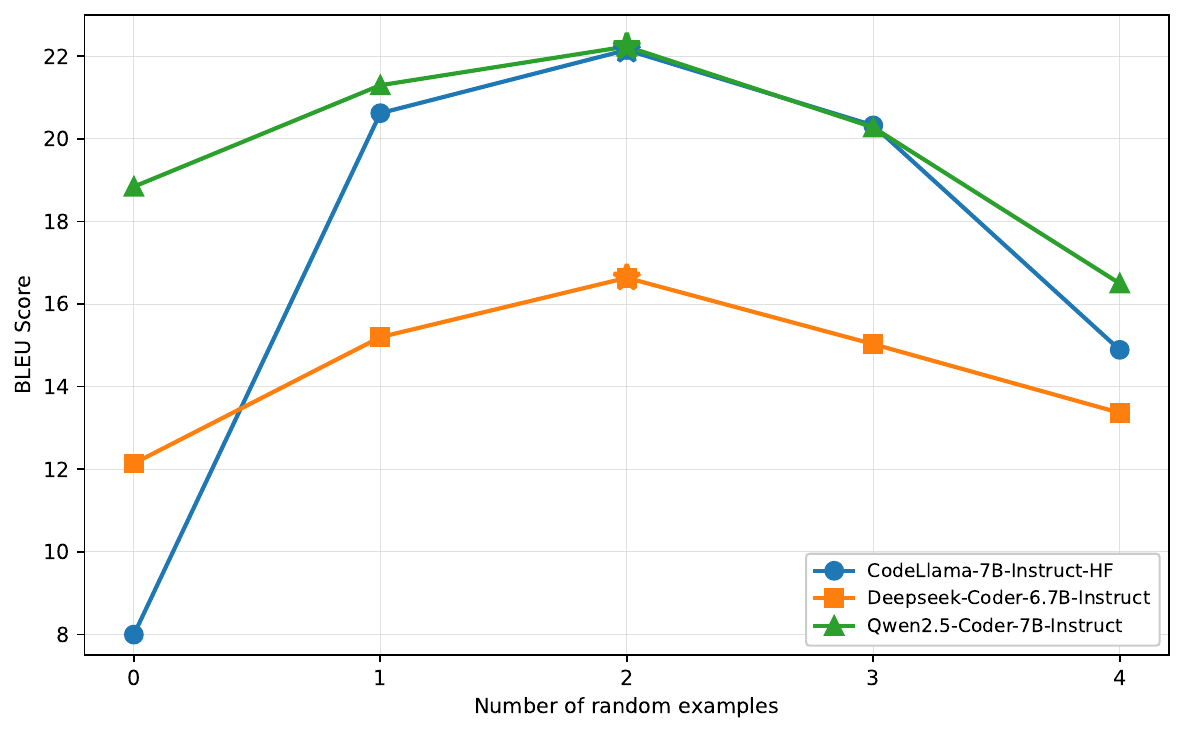}
  }
  \hfill 
  \subfigure[SolidityScore]{
    \label{fig:icl-solidity}
    \includegraphics[width=0.47\linewidth]{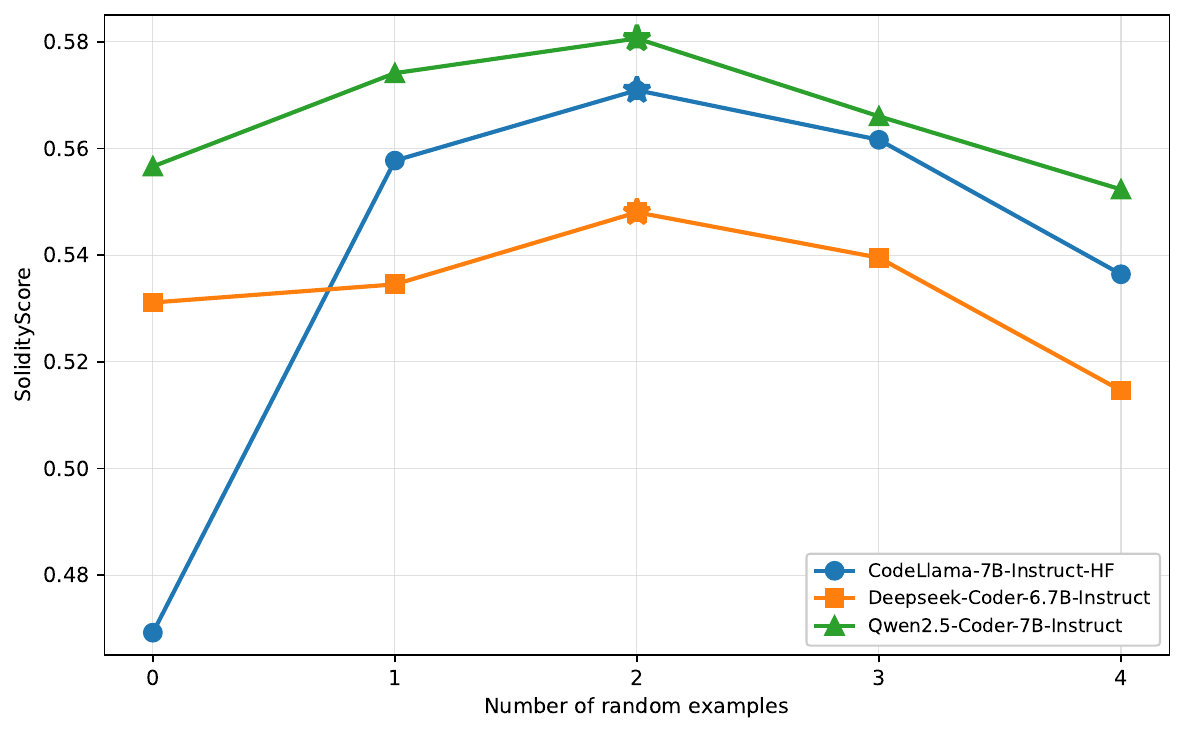}
  }
  
  \caption{Performance trends of evaluated models using ICL with varying numbers of demonstrations ($k \in \{0, 1, 2, 3, 4\}$).}
  \label{fig:fig4-icl_result}
\end{figure*}

\noindent\textbf{Results.}
The experimental results are shown in Fig. \ref{fig:fig4-icl_result}. Taking Qwen2.5-Coder as an example, its performance on syntactic accuracy (BLEU) and semantic consistency (SolidityScore) improved with the addition of in-context examples. Starting with a baseline BLEU score of 18.8433 in the zero-shot setting, the score rose to 21.2989 with one example and peaked at 22.2377 with two examples. Similarly, the SolidityScore increased from a baseline of 0.5566 to a peak of 0.5806 at the two-shot setting, indicating that the model captured deeper logical semantics. However, this upward trend reversed when more than two examples were provided, with both metrics declining at the three- and four-shot settings

The other two models exhibited the same pattern of initial improvement followed by decline. CodeLlama showed the most dramatic gains: its BLEU score jumped from 7.9957 in the zero-shot setting to a peak of 22.1527 with two examples, while its SolidityScore rose correspondingly from 0.4692 to 0.5709. DeepSeek-Coder followed a similar trend, with BLEU improving from 12.1412 to 16.6359 and SolidityScore from 0.5311 to 0.5480 at the two-shot setting. However, as shown in Fig. \ref{fig:fig4-icl_result}, DeepSeek-Coder's performance deteriorated most sharply when the context expanded beyond two examples. These results collectively demonstrate that a limited number of in-context examples effectively activates domain knowledge, whereas an excessive context size leads to performance degradation across all models.


We hypothesize that this phenomenon is likely attributable to the inherent complexity and high token consumption characteristic of repository-level tasks. In contrast to function-level code generation tasks, repository-level synthesis entails extensive logic and cross-file dependencies, meaning that each demonstration imposes a significant burden on the limited context window. Consequently, stacking multiple repository-level examples ($k \geq 3$) results in an excessively long context. This overwhelms the model's attention mechanism, introducing conflicting coding styles and irrelevant details as noise. Instead of providing useful information, this information overload distracts the model from the target requirement. Therefore, providing exactly two high-quality examples offers the best balance, avoiding the negative impact of long-context saturation.

\noindent\textbf{Answer to RQ2.2.}
Our empirical analysis reveals a consistent saturation point for in-context learning in repository-level Solidity code generation, where performance consistently peaks at two examples ($k=2$) across all evaluated models. This suggests that a small number of examples is sufficient to convey essential structural and stylistic patterns of Solidity code. Beyond this point, adding more demonstrations leads to diminishing returns and, in some cases, measurable performance degradation.

\subsubsection{RQ2.3 Effectiveness of Retrieval-Augmented Generation}\mbox{}\\
\leavevmode\noindent \textbf{Methodology.}
To address the limitations of static ICL, where a fixed set of randomly selected examples is reused for all test queries regardless of their content, as in RQ2.2, we investigate RAG as a dynamic alternative. RAG selects query-specific examples at inference time, mitigating the risk that randomly chosen demonstrations may be irrelevant to a given task. We implement a RAG framework in which the entire training set (4,376 high-quality samples) is indexed as an external knowledge base. During inference, we apply the BM25 algorithm~\cite{RobertsonZ09} to compute textual similarity between the natural language problem description of a test query and samples in the knowledge base. For each query, the top-\(k\) most similar examples (\(k \in \{1, 2, 3, 4\}\)) are retrieved and incorporated into the prompt. This dynamic retrieval strategy helps ensure that the model conditions on code examples with closely related logic and functional requirements, providing more targeted domain guidance than generic few-shot demonstrations.

\begin{figure*}[!htbp] 
  \centering 
  \subfigure[BLEU]{
    \label{fig:rag-bleu}
    \includegraphics[width=0.47\linewidth]{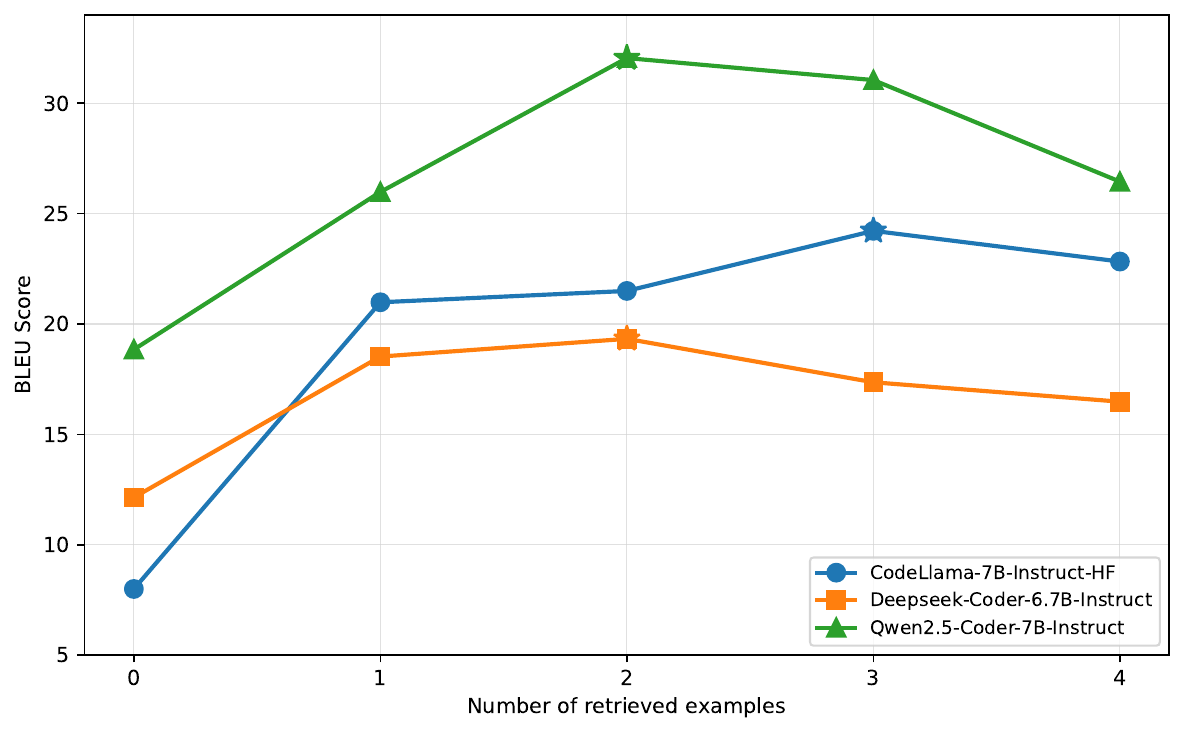} 
  }
  \hfill 
  \subfigure[SolidityScore]{
    \label{fig:rag-solidity}
    \includegraphics[width=0.47\linewidth]{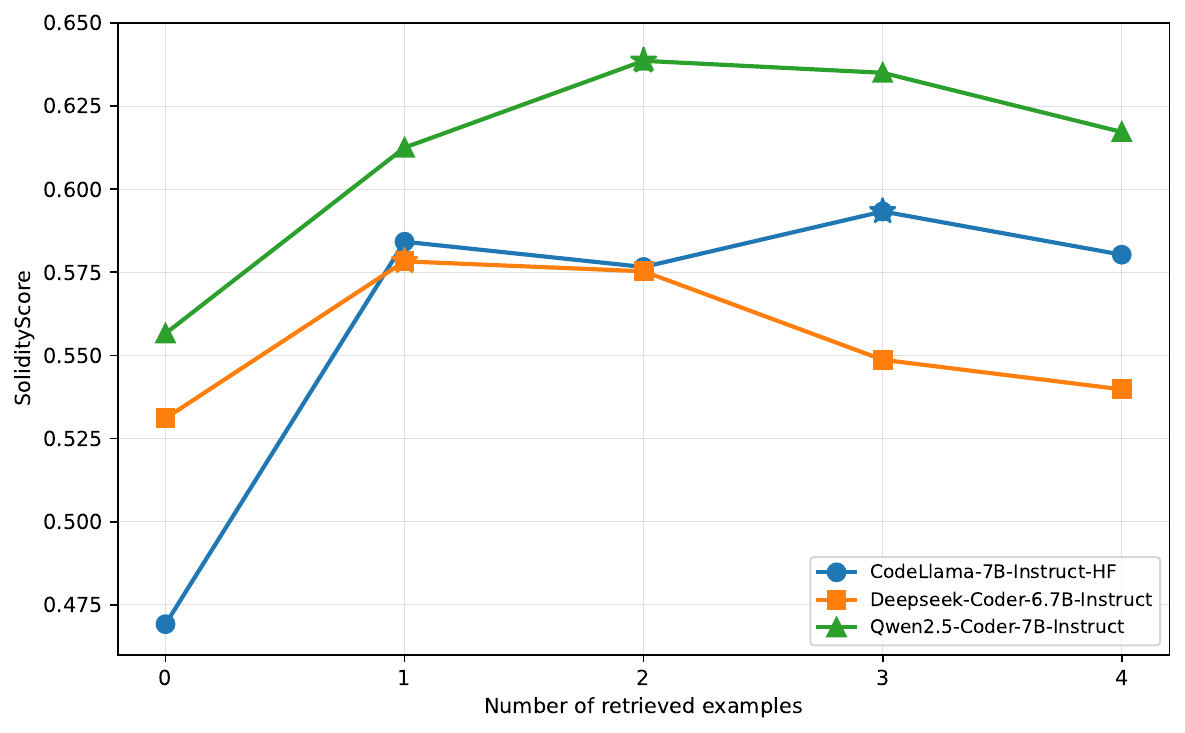}
  }  
  \caption{Performance trends of evaluated models using RAG with varying numbers of demonstrations ($k \in \{0, 1, 2, 3, 4\}$).} 
  \label{fig:fig5-rag_result} 
\end{figure*}

\noindent\textbf{Results.}
As illustrated in Fig. \ref{fig:fig5-rag_result}, the RAG strategy significantly bolstered the code generation performance of all evaluated models compared to ICL. All models showed a consistent pattern: performance initially increased with the number of retrieved samples, then declined as the context expanded further.

Qwen2.5-Coder demonstrated the most substantial gain. Its BLEU score rose from 25.9813 ($k=1$) to a peak of 32.0489 at two retrieved samples ($k=2$), while its SolidityScore simultaneously increased from 0.6125 to 0.6386. Notably, this peak performance represents a 44.1\% relative improvement in BLEU and a 10.0\% improvement in SolidityScore compared to its best static ICL result. DeepSeek-Coder followed a similar trend, peaking at $k=2$ with a BLEU score of 19.3276 and a SolidityScore of 0.5753. This corresponds to a 16.1\% relative improvement in BLEU over the static ICL baseline. 
In contrast, CodeLlama exhibited a delayed peak. Its BLEU score continued to rise from 20.9843 ($k=1$) to reach its zenith of 24.22 at three retrieved samples ($k=3$), while its SolidityScore rose from 0.5842 to 0.5933. This represents a 9.3\% relative improvement in BLEU and a 3.9\% improvement in SolidityScore over its optimal ICL setting. It was only when the retrieved samples increased to four that its performance began to decline, indicating a distinct tolerance for context length compared to the other models.

Our analysis of the performance trends and model-specific variations (Fig. \ref{fig:fig5-rag_result})  identifies two key mechanisms driving these results. First, RAG outperforms static ICL because it retrieves semantically relevant code that directly maps to the target requirements. However, the observed rise and fall trend confirms that context saturation is inevitable; while 2-3 relevant examples provide crucial logic and security patterns, stacking more examples ($k \ge 4$) creates an excessively long context. This introduces information redundancy and cognitive noise that distracts the model from the core task, offsetting the benefits of retrieval. Second, the observed divergence in optimal retrieval volume—where Qwen2.5 peaks at $k=2$ while CodeLlama peaks at $k=3$, suggests a fundamental difference in their in-context learning efficiency. Newer models like Qwen2.5 appear capable of extracting sufficient task patterns from minimal context; thus, two high-quality examples provide ample guidance, rendering additional examples redundant. In contrast, models with lower baselines like CodeLlama seem to benefit from extended context, where a third example likely reinforces the generation paradigm. However, beyond these peaks, the increased context length introduces information overload that outweighs any marginal benefit, ultimately leading to performance degradation.

\noindent\textbf{Answer to RQ2.3.} RAG consistently outperforms static ICL, which relies on randomly sampled examples, across all evaluated models, highlighting the benefit of dynamically retrieving query-relevant demonstrations. Performance typically peaks with a small number of retrieved examples and degrades as more are added, indicating context saturation. The optimal retrieval size varies across models, suggesting differences in their in-context learning efficiency.


\subsection{RQ3: Impact of SFT on Domain Adaptation}

\leavevmode\noindent \textbf{Methodology.} 
To investigate the role of parameter updates in achieving deep domain adaptation, we performed SFT on our constructed high-quality Solidity instruction dataset. Given the computational costs associated with full-parameter tuning, we adopted LoRA~\cite{DBLP:SFT0} as a parameter-efficient strategy. The configuration involved a rank $r=16$ and a scaling factor $\alpha=32$. To maximize the model's representational capability, LoRA adapters were injected into all critical linear layers of the Transformer architecture, including the query ($q\_proj$), key ($k\_proj$), value ($v\_proj$), and output ($o\_proj$) projection layers, as well as the gate, up, and down projection layers within the feed-forward networks.

The training process utilized 8-bit floating point (FP8) to quantization to optimize memory efficiency. All samples were formatted into a standard instruction-following structure. We employed a standard language modeling loss with masking as the optimization objective, calculating loss only on the code generation tokens while ignoring the instruction prefix. Training hyperparameters included  a learning rate of $3e^{-4}$, using a cosine annealing scheduler with a 5\% warmup ratio to ensure convergence and prevent overfitting.

\noindent\textbf{Results.}
The experimental results presented in Table \ref{tab:finetuning_performance} demonstrate that SFT triggered a fundamental improvement in model performance, significantly surpassing all non-parametric prompt engineering strategies (ICL and RAG). All fine-tuned models achieved a qualitative breakthrough, with performance metrics reaching new heights across the board.

\begin{table}[!htbp]
    \centering
    \caption{Performance of models in the SFT setting}
    \label{tab:finetuning_performance}
    \begin{tabular}{lcc}
        \toprule
        \textbf{Model} & \textbf{BLEU} & \textbf{SolidityScore} \\
        \midrule
        CodeLlama-7B-Instruct-HF     & 28.1663 & 0.6361 \\
        Deepseek-Coder-6.7B-Instruct & 25.3060 & 0.6240 \\
        Qwen2.5-Coder-7B-Instruct    & \textbf{35.9584} & \textbf{0.6465} \\
        \bottomrule
    \end{tabular}
\end{table} 

Qwen2.5-Coder-7B-Instruct exhibited exceptional domain adaptability. Its BLEU score soared to 35.96, and its SolidityScore reached 0.6465, establishing absolute dominance in the comparative experiments. This represents a significant improvement over its best RAG performance (BLEU: 32.05). CodeLlama-7b-Instruct-hf also recorded substantial gains. Its BLEU score reached 28.17, while its SolidityScore rose to 0.6361. Notably, this post-SFT performance significantly outperforms its RAG peak (BLEU: 24.22), proving that parameter optimization effectively bridges the gap between older architectures and domain-specific requirements. DeepSeek-Coder-6.7b-Instruct similarly showed robust growth, attaining a BLEU score of 25.3060 and a SolidityScore of 0.6240. The fact that the SolidityScore for all models exceeded the 0.62 threshold indicates that SFT enables models to master not just the syntax, but the deeper semantic logic of smart contracts.

The experimental results reveal two key factors underlying the superior performance of SFT in domain adaptation. First, SFT’s ability to internalize domain knowledge proves more effective than external guidance mechanisms such as ICL and RAG. While ICL and RAG depend on retrieved context to ``remind'' the model of relevant knowledge--an approach constrained by context length and retrieval quality, SFT directly encodes domain-specific patterns into the model’s parameters. This eliminates reliance on noisy or incomplete external prompts and enables the model to absorb strict domain constraints (e.g., security rules and gas optimization in Solidity) as intrinsic capabilities. Consequently, SFT yields more consistent outputs and lower inference latency.

Second, SFT unlocks significant potential across varied model architectures. Although newer, more capable models (e.g., Qwen2.5) attain higher absolute scores due to stronger pre-training, older models (e.g., CodeLlama) exhibit a steeper ``adaptation curve'' when fine-tuned with high-quality domain data. The dramatic improvement observed in CodeLlama indicates that even models with initially weaker instruction-following abilities can be transformed into specialized experts through targeted SFT. This underscores that, for technical domains like Solidity, a well-pre-trained base model coupled with a high-quality SFT pipeline is critical to achieving industrial-grade reliability.

\noindent\textbf{Answer to RQ3.} 
SFT is a decisive factor for effective domain adaptation in code generation, particularly in specialized, constraint-rich domains like Solidity. SFT surpasses RAG or ICL methods by internalizing domain knowledge directly into the model parameters, thereby reducing inconsistency and latency. Moreover, SFT enables substantial performance gains across diverse model architectures, including older or less instruction-tuned models, when applied with high-quality, domain-specific data. Thus, for industrial-strength domain adaptation, a robust SFT pipeline built upon a sufficiently pre-trained base model is essential.


\subsection{RQ4: SolidityScore Reliability Verification} \label{sec:rq4}


\noindent \textbf{Methodology.} To validate the reliability of SolidityScore, particularly its ability to correctly assess functionally equivalent but structurally distinct code, we designed an adversarial experiment. This addresses a key limitation of traditional $N$-gram metrics like BLEU, which are overly sensitive to superficial lexical features and often penalize functionally correct code merely for differences in coding style or syntax. For the experiment, we constructed an adversarial test set of 100 high-quality samples. Critically, instead of applying simple rule-based code transformations, we leveraged the GPT-4 API as a semantic-preserving perturbation engine. By using meticulously designed prompt templates, we instructed GPT-4 to generate diverse code variants that preserve the original functionality, thereby creating a rigorous test for metric robustness.

Specifically, while strictly constraining the business logic and control flow to remain unchanged, we performed multi-dimensional refactoring of the code. At the lexical level, we implemented aggressive variable anonymization, such as replacing descriptive identifiers (e.g., \texttt{userBalance}) with generic tokens like \texttt{v1}. Structurally, we applied equivalent rewrites to loop logic and conditional branches, such as converting for loops into while structures or reconfiguring the nesting of if-else statements. This process generated adversarial samples that remain functionally identical to the originals despite exhibiting extremely low textual similarity. The objective was to force the evaluation metrics to penetrate surface-level noise and capture the underlying core semantics. We compared the stability of BLEU and SolidityScore on 50 semantically equivalent but syntactically perturbed adversarial samples.

\begin{figure*}[!htbp]
  \centering
  \includegraphics[width=0.5\linewidth]{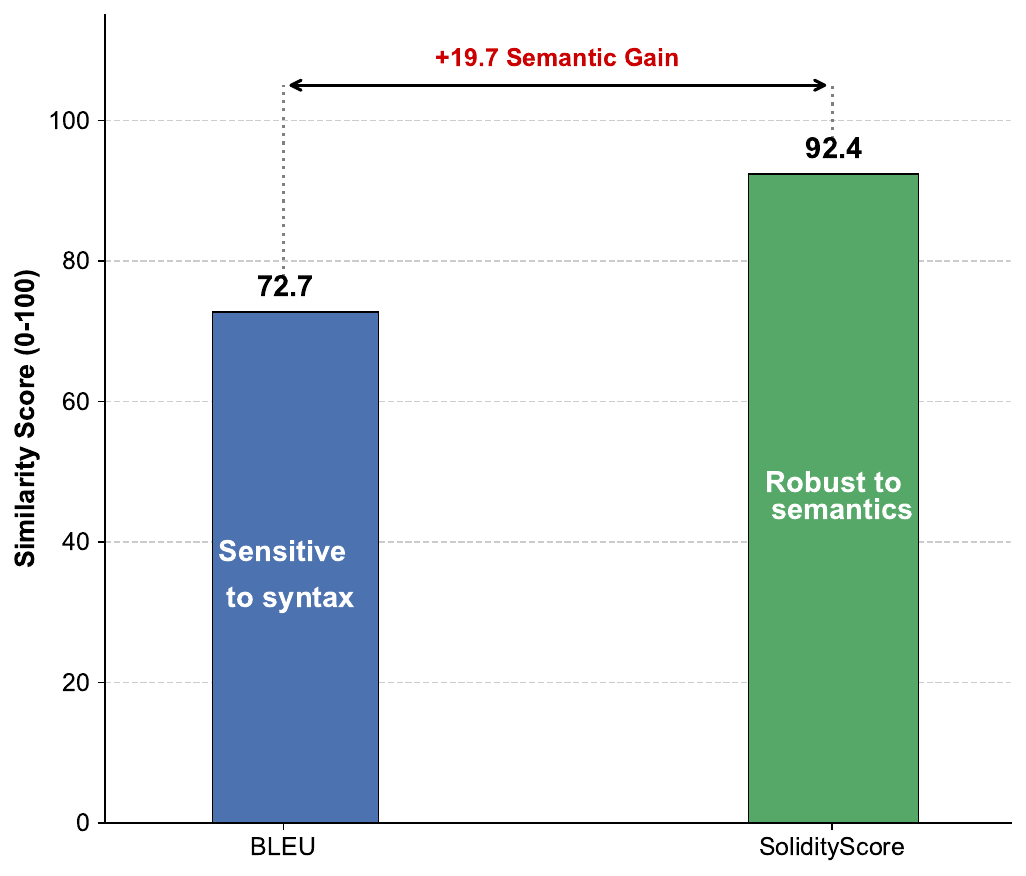}
  \caption{Evaluation of metric robustness under adversarial attacks}
  \label{fig:fig7-robustness_comparison}
\end{figure*}

\noindent \textbf{Results.}  The experimental results, shown in Fig. \ref{fig:fig7-robustness_comparison}, reveal a fundamental difference in their performance under adversarial conditions. Constrained by its rigid dependency on surface-level tokens, the BLEU metric proved fragile when faced with variable renaming and structural rewriting, achieving an average score of only 72.7. This outcome indicates that evaluation methods based on textual overlap struggle to accommodate the diversified output styles of generative models, posing a significant risk of quality underestimation. In contrast, SolidityScore demonstrated exceptional semantic stability, maintaining a high average score of 92.4. Quantitative analysis shows that this corresponds to a semantic gain of +19.7 over the baseline BLEU score. This pronounced performance advantage can be attributed to its underlying Solidity-LLM encoder, which underwent large-scale domain-specific instruction fine-tuning to map code into a high-dimensional semantic space.

\noindent\textbf{Answer to RQ4.} SolidityScore effectively filtered out perturbations in variable names and syntactic structures while accurately identifying the inherent logical consistency of the code. Therefore, this experiment provides compelling evidence that SolidityScore possesses superior construct validity. It offers a more objective and robust quality measure for smart contract generation tasks compared to traditional lexical overlap metrics  (e.g., BLEU).

\subsection{RQ5: Compilability Analysis of SFT-generated Code across Different Models} 

\noindent \textbf{Methodology.} While metrics like BLEU and SolidityScore provide complementary views, respectively assessing surface-form similarity and deep semantic equivalence, they share a fundamental limitation: neither can guarantee the practical executability of generated code. A contract may achieve high scores on these metrics yet fail to compile due to unresolved dependencies, version mismatches, or project-level configuration errors. To examine the compilability performance of code generated under the SFT paradigm, we transitioned from static metric-based assessment to a compiler-based validation paradigm. Given the negligible compilation success of prompt-based methods (CoT, ICL, and RAG) in preliminary repository-level tasks, we focus this assessment exclusively on the SFT paradigm. Specifically, we employ two metrics: the Compilation Success Rate (CSR), which measures the ratio of samples that compile without errors using the Solidity compiler (\texttt{solc}) version \texttt{0.8.20}, and the Syntax Correctness Rate (SCR), which measures samples that can be parsed into a valid AST but may fail compilation due to external dependencies. 

Additionally, an automated parsing pipeline was implemented to categorize every compilation failure based on compiler feedback derived from the \texttt{solc}, resulting in a detailed taxonomy of error types (e.g., missing context, missing dependency, Markdown/Text noise). This methodology allows us to quantify compilability and diagnose the root causes of failure in a repository-level context.

\begin{figure*}[!htbp]
  \centering
   \includegraphics[width=0.6\linewidth]{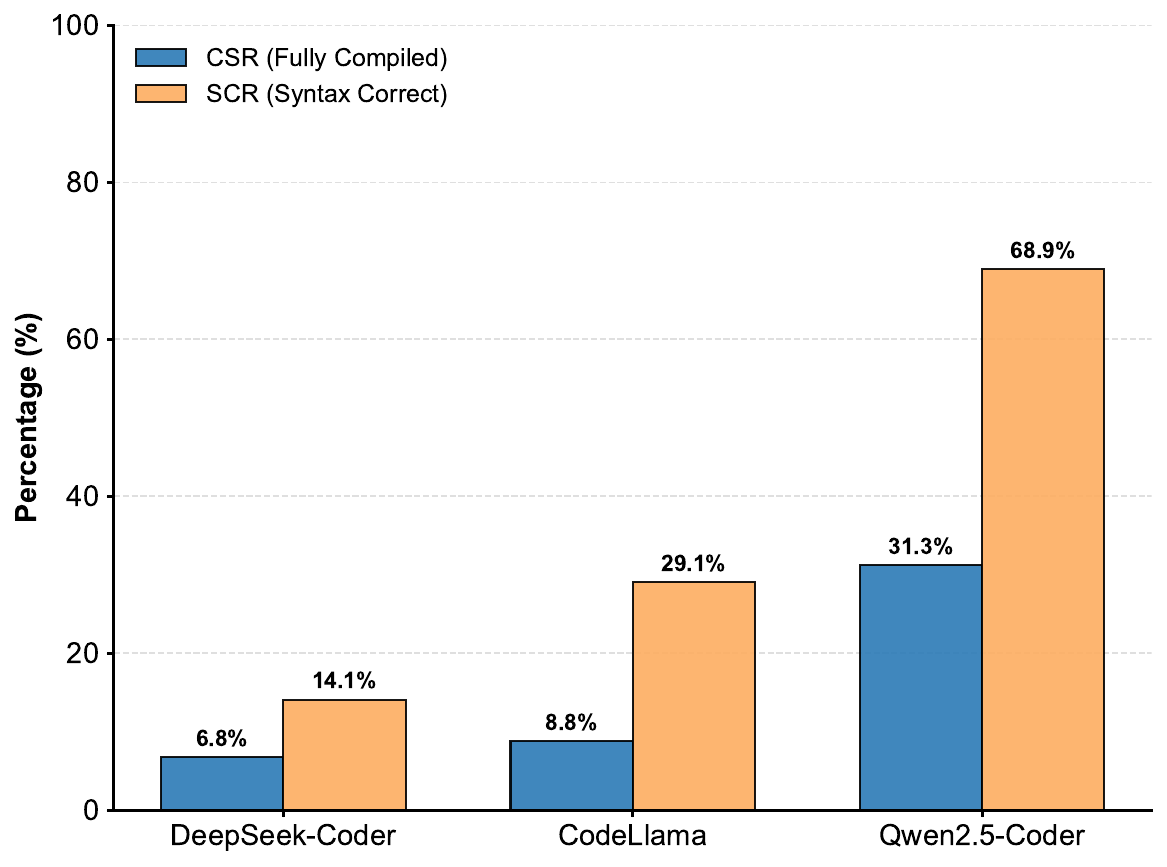}
   \caption{Comparison of compilability for code generated by different LLMs in the SFT setting. The significant gap between SCR and CSR indicates the challenge of meeting strict Solidity compiler constraints beyond basic syntax.}
  \label{fig:fig7-compile_metrics}
\end{figure*}

\noindent \textbf{Results.} The compilablity results for code generated by different LLMs in the SFT setting are shown in Fig. \ref{fig:fig7-compile_metrics}. Qwen2.5-Coder significantly outperformed CodeLlama and DeepSeek-Coder across both metrics, achieving an SCR of 68.9\% and a CSR of 31.3\%, whereas the CSR for the latter two remained below 10\%, at 8.8\% and 6.8\% respectively. Despite the overall performance variance, all models exhibited a substantial gap between SCR and CSR, suggesting that a significant portion of syntactically parsable code remains unusable at an engineering level.



 This observed discrepancy between high syntax correctness and low compilation success rate stems primarily from the lack of repository-level context in conventional instruction tuning datasets. By learning from high-quality, domain-specific samples, SFT enables models to closely approximate the distribution of Solidity code at the statement level. This allows them to master constructs such as function definitions, visibility modifiers, and event declarations, which significantly improves the SCR. The successful compilation of a Solidity contract also depends on external dependencies, such as library import paths, compiler version pragmas, and inheritance relationships across multiple files. As a result, while models can generate syntactically correct code fragments, they often fail to resolve these cross-file references correctly, leading to linking errors in practical compilation environments. This limitation highlights the current ceiling of SFT for generating fully compilable, repository-level smart contracts. 

\begin{figure*}[!htbp]
  \centering
   \includegraphics[width=0.8\linewidth]{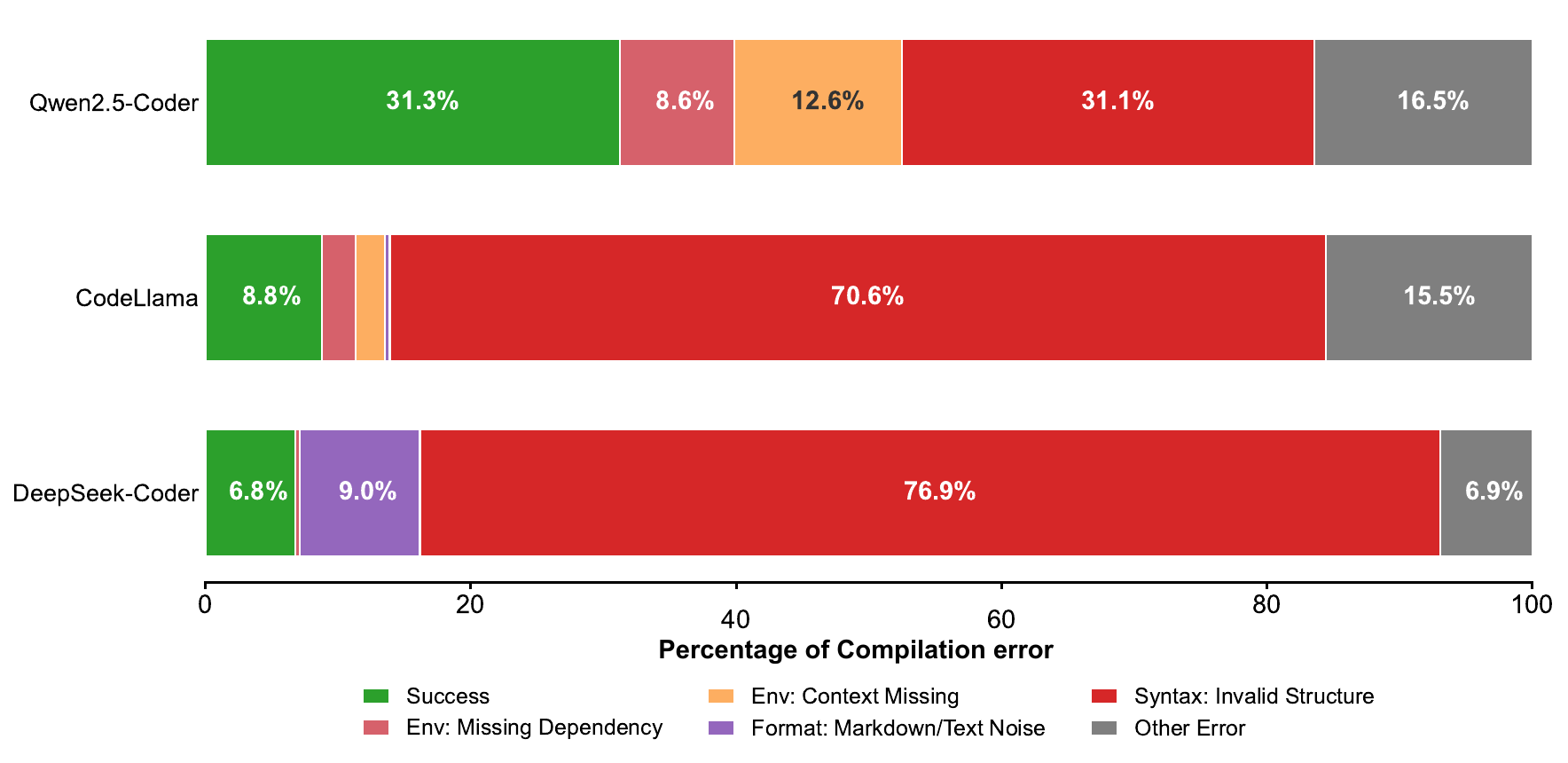}
  \caption{Analysis of Compilation Failure Causes. The chart distributes compilation errors based on compiler feedback, categorizing failures into types like syntax errors, missing dependencies, and format noise. The result highlights a model capability gap: less capable models are limited by syntax, while advanced models like Qwen2.5-Coder are challenged by project-level environmental context.}
  \label{fig:fig8-compile_taxonomy}
  \end{figure*}

 The resulting taxonomy of compilation failures is presented in Fig. \ref{fig:fig8-compile_taxonomy}. For CodeLlama and DeepSeek-Coder, most failures stemmed from syntax errors, constituting 70.6\% and 76.9\% of their respective unsuccessful samples. This suggests that a significant portion of generated contracts is rejected during the initial parsing phase, often due to structurally incomplete definitions or invalid modifier usage. Moreover, DeepSeek-Coder displayed a 9.0\% error rate attributed to ``Format: Markdown/Text Noise'', indicating some instability in adhering strictly to code-generation instructions. These observations imply that when a base model’s pre-training coverage of a strongly typed, domain-specific languages like Solidity is inherently limited, downstream instruction tuning alone cannot fundamentally reshape its underlying syntax generation patterns.

In contrast, Qwen2.5-Coder displayed a distinct error profile; its structural error rate dropped to 31.1\%, while a higher proportion of failures stemmed from environment-related categories such as ``Env: Missing Dependency'' (8.6\%) and ``Env: Context Missing'' (12.6\%). This shift suggests that given sufficient syntactic knowledge, SFT encourages models to attempt more complex patterns involving external libraries, which manifests as environmental hallucinations in the absence of a real project structure. Overall, while SFT effectively reduces low-level syntactic errors in high-quality base models, it remains insufficient for ensuring the stable generation of compilable project-level Solidity contracts without integrated build environment information.

\noindent\textbf{Answer to RQ5.}  The compiler-based evaluation reveals that while LLMs in the SFT setting can generate a substantial portion of syntactically valid Solidity code (e.g., achieving a SCR of up to 68.9\%), their ability to produce fully compilable, repository-level contracts remains severely limited, with the highest CSR reaching only 31.3\%, and performance across models varies significantly, with the CSR for others falling below 10\%. Critically, although the specific distribution of error subtypes differs between models, ranging from fundamental grammatical mistakes to incorrect project-level references (e.g., missing dependencies), the evidence indicates that syntax errors constitute the principal cause of compilation failure.
\section{Case Study}
\label{chap:chap6}

While quantitative experiments have validated the significant performance gains of SFT models over their zero-shot baselines, we conducted this qualitative case study on a representative bidding refund scenario to explore how instruction tuning fundamentally transforms the model's approach to logic construction and architectural robustness. Unlike prompt-engineering strategies (such as CoT, ICL, and RAG) that rely on external guidance to mitigate structural flaws, SFT internalizes complex programming constraints directly into the model's parameters. The purpose of this dedicated analysis, moving beyond the brief qualitative notes in RQs, is to granularly expose the reasoning gap in a complex business logic: a bidding refund system.

\begin{figure*}[!htbp]
    \centering
    \setlength{\tabcolsep}{1pt}
    \subfigure[Ground Truth]{
        \includegraphics[width=0.45\linewidth]{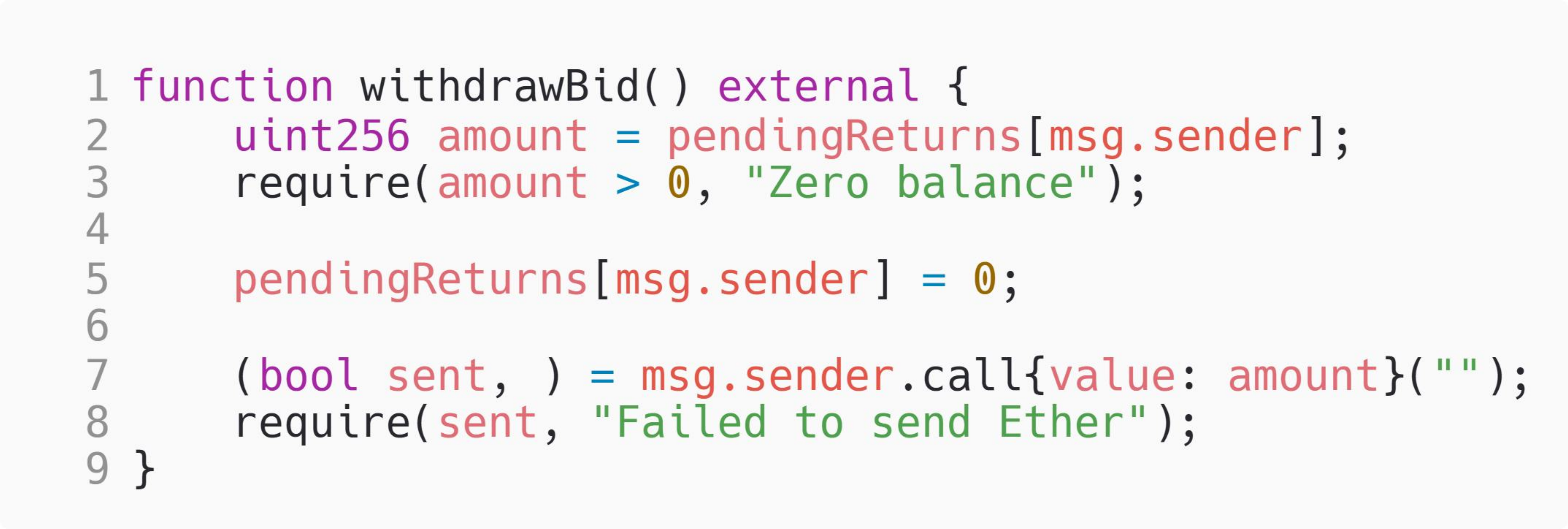}
        \label{fig:case_gt}
    }
    \hfill
    \subfigure[Zero-shot]{
        \includegraphics[width=0.45\linewidth]{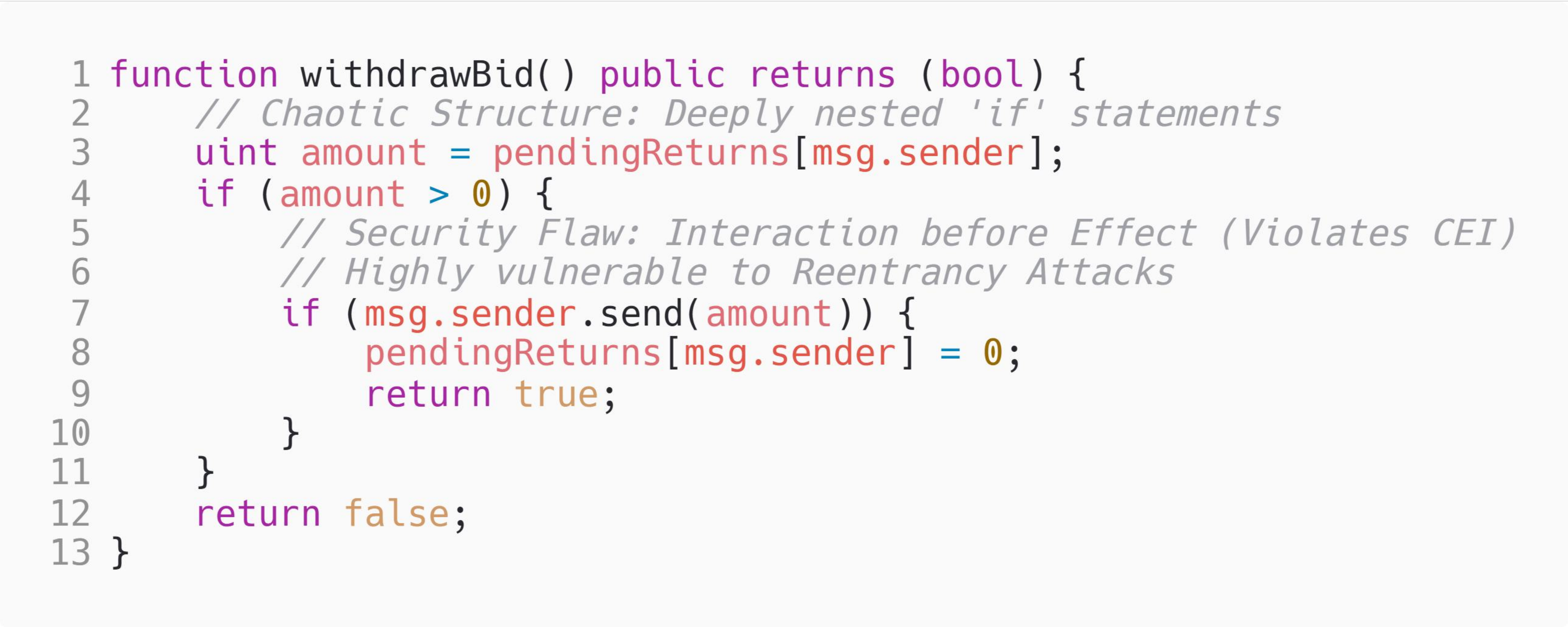}
        \label{fig:case_zs}
    }
    \vspace{0.1cm}
    \subfigure[CoT]{
        \includegraphics[width=0.45\linewidth]{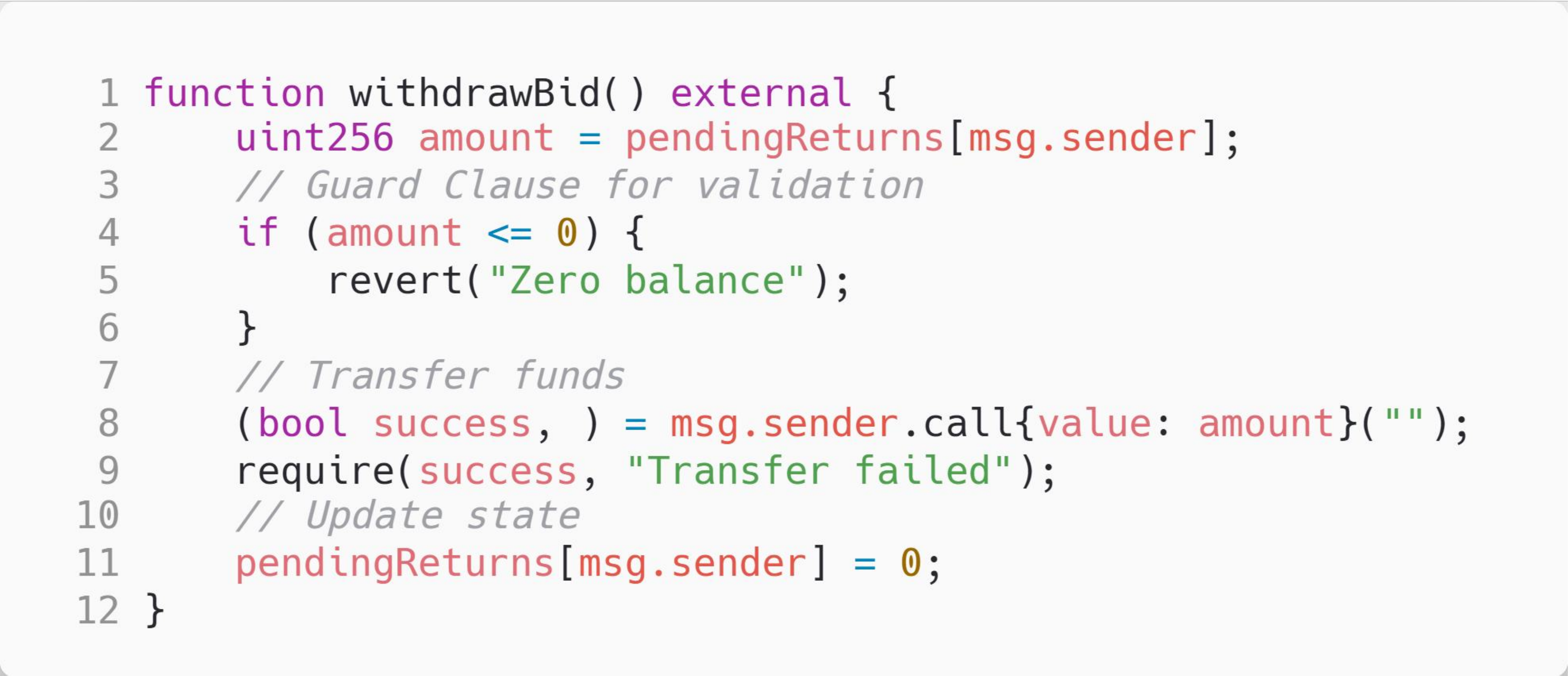}
        \label{fig:case_cot}
    }
    \hfill
    \subfigure[ICL]{
        \includegraphics[width=0.48\linewidth]{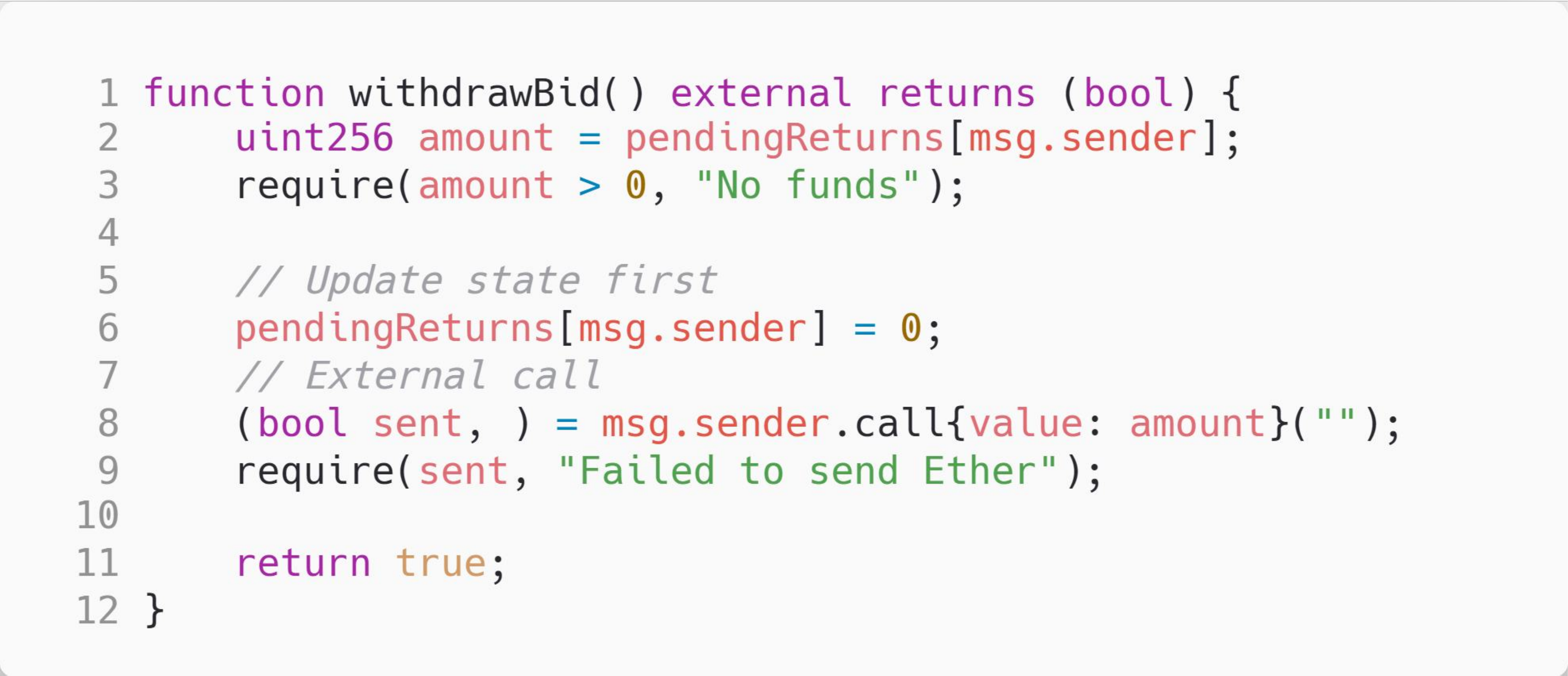}
        \label{fig:case_icl}
    }
    \vspace{0.1cm}
    \subfigure[RAG]{
        \includegraphics[width=0.45\linewidth]{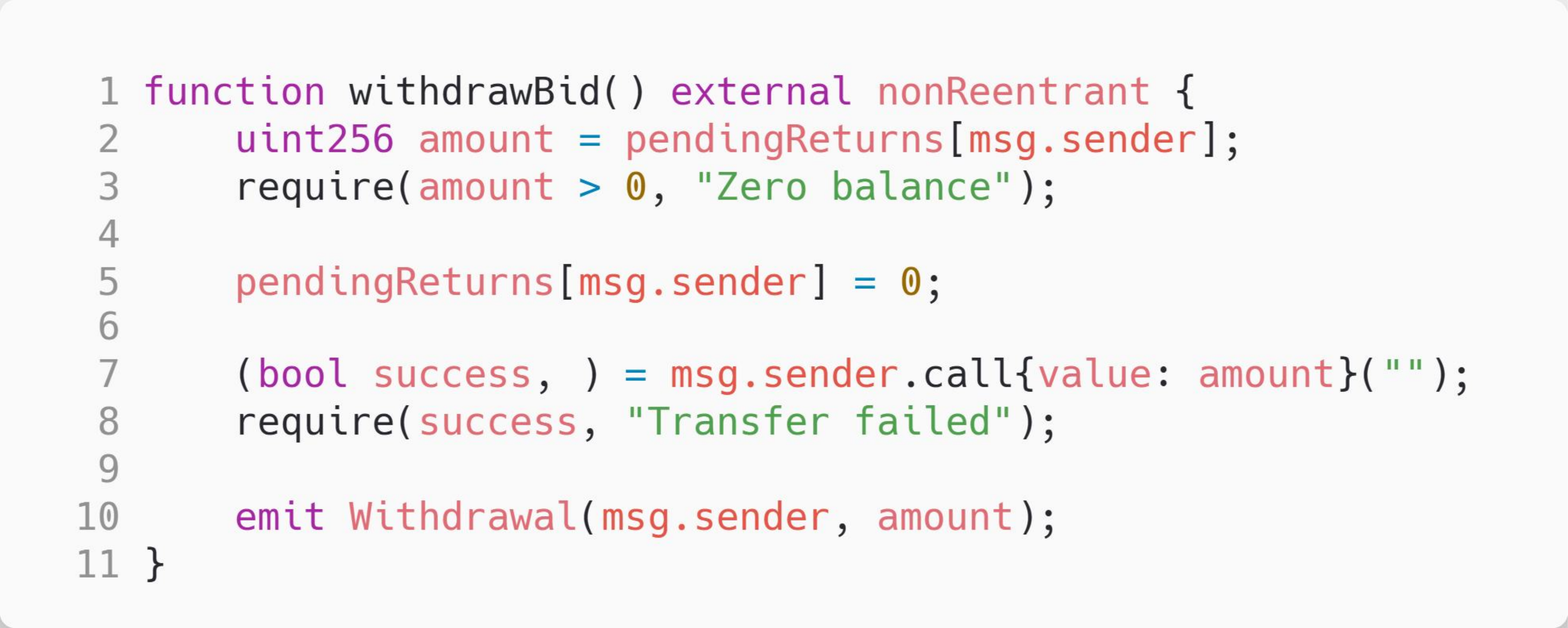}
        \label{fig:case_rag}
    }
    \hfill
    \subfigure[SFT]{
        \includegraphics[width=0.45\linewidth]{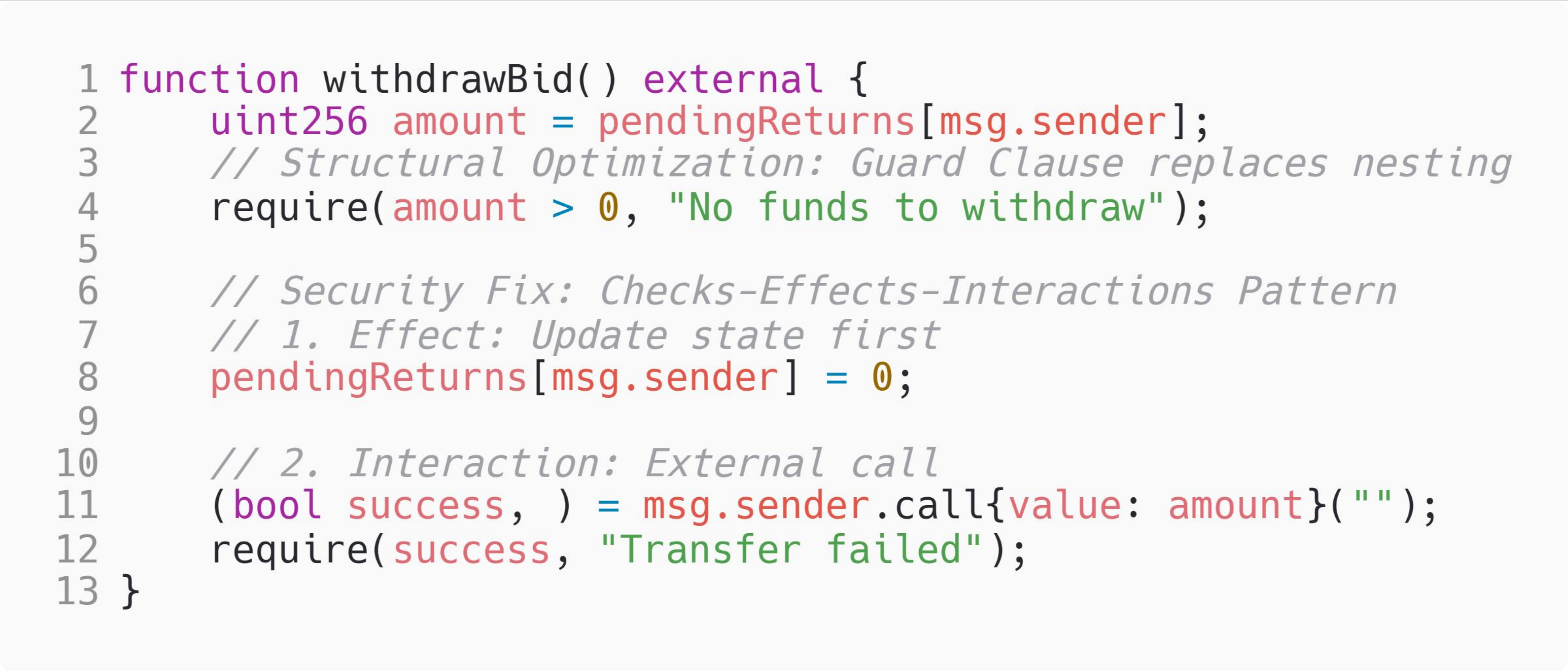}
        \label{fig:case_sft}
    }
    \caption{Qualitative comparison among different adaptation paradigms for the ``Auction Withdrawal'' task, based on outputs from Qwen2.5-Coder-7B-Instruct. 
    (a) \textbf{Ground Truth} serves as the secure reference implementation. 
    (b) \textbf{Zero-shot} exhibits chaotic nesting and a critical re-entrancy vulnerability (Interaction-before-Effect). 
    (c) \textbf{CoT} improves structural clarity via guard clauses but fails to resolve the logical vulnerability. 
    (d) \textbf{ICL} adopts the CEI security pattern through example imitation. 
    (e) \textbf{RAG} introduces domain-specific protection via knowledge retrieval. 
    (f) \textbf{SFT} achieves the optimal synthesis of structural optimization and strict adherence to security standards.}
    \label{fig:fig6-case_analyse}
\end{figure*}

As illustrated in Fig. \ref{fig:fig6-case_analyse}, the adaptation paradigms exhibited substantially different capability boundaries when addressing this task using  Qwen2.5-Coder-7B-Instruct. The code synthesized by the pre-trained model in a zero-shot setting (Fig. \ref{fig:fig6-case_analyse} (b)) displayed significant structural defects and security blind spots. Structurally, it relied on deeply nested conditional statements—an ``arrow-shaped'' architecture that increases the cognitive load for auditors and obscures the core business logic. More critically, the model operating in the zero-shot setting severely violated the CEI pattern, a cornerstone of secure smart contract development. The code erroneously placed external transfer operations before state updates, sending funds before deducting the user's balance. This ``interaction-before-effect'' logical flaw renders the contract highly vulnerable to re-entrancy attacks, where an attacker could utilize a malicious contract to recursively call the function before the balance is zeroed, thereby exhausting the contract's funds.

As shown in Fig. \ref{fig:fig6-case_analyse}(c),  the CoT strategy addressed these structural deficiencies that arise in the zero-shot setting by guiding the model through explicit reasoning steps. While this approach successfully optimized the control flow by replacing deep nesting with guard clauses, thereby making the code easier to read and follow, the logical reasoning failed to rectify the underlying domain misconception. The CoT-generated code retained the critical ``interaction-before-effect'' flaw, executing the external call prior to the state update, which leaves the re-entrancy vulnerability unresolved despite the improved readability. 
Conversely, the ICL paradigm (Fig. \ref{fig:fig6-case_analyse}(d)) effectively regularized the generative style through pattern imitation. By attending to high-quality demonstrations, the model correctly implemented the CEI pattern. However, this reliance on imitation often leads to mechanical redundancy. As shown in Fig. 10(d), the model includes an unnecessary return true; statement, reflecting its inability to distinguish core logic from the decorative artifacts of the provided examples.
Unlike the above strategies, RAG injected external domain knowledge to supplement the model's internal representations (Fig. \ref{fig:fig6-case_analyse}(e)). By retrieving relevant logic, the model incorporated advanced defense patterns like the nonReentrant modifier. Nevertheless, the output exhibits an over-reliance on external components. The inclusion of the emit statement and \texttt{nonReentrant} modifier—while functional—introduces unnecessary complexity and potential compilation risks if specific library dependencies are absent in the project environment.

The fine-tuned model (Fig. \ref{fig:fig6-case_analyse}(f)) distinguishes itself through structural optimization and logical autonomy. As illustrated in the code, SFT achieves a highly refined implementation that mirrors the Ground Truth without the redundant return statements found in ICL or the external modifier dependencies introduced by RAG. By autonomously adopting a flattened architecture with guard clauses, SFT replaces the chaotic nesting observed in the zero-shot baseline. More importantly, it demonstrates a sophisticated use of modern Solidity syntax, specifically the \texttt{.call} method, ensuring the generated code is not only correct but also aligned with current engineering best practices. This superior performance stems from training on our meticulously curated high-quality corpus, which enables the model to master Solidity's core syntactic logic and specialized structures—such as the CEI pattern. By internalizing this specialized Solidity knowledge directly into the model's parameters, SFT significantly enhances the model's capability in Solidity code generation, allowing it to synthesize robust and optimized solutions without relying on any external cues.


Additionally, to objectively validate our manual analysis, we employed Slither \cite{5.6:Slither/FeistGG19}, a static analysis tool for Solidity smart contracts, to check for the presence of vulnerabilities. It can detect typical high-risk vulnerabilities such as  integer overflow/underflow and reentrancy. The automated detection results confirmed our findings. For the zero-shot and CoT outputs, Slither explicitly flagged a re-entrancy vulnerability, reporting the error ``State variables written after the call(s)'' and pinpointing that \texttt{pendingReturns} was updated subsequent to the external \texttt{send} interaction. In contrast, the ICL, RAG, and SFT-generated codes successfully passed the re-entrancy check. Specifically, ICL passed by mimicking the CEI pattern, while RAG passed by applying the re-entrancy guard. This objective verification highlights that while prompting strategies (ICL/RAG) can effectively mitigate risks through external guidance, SFT achieves robust security by internalizing critical constraints, ensuring valid generation even in the absence of external cues.

\section{Discussion}
\label{chap:chap7}
\subsection{Implications}
According to our empirical findings, we summarize three key future research directions for domain-specific code generation: the evolution of evaluation frameworks, the advancement of optimization methodologies, and the expansion of data foundations.

First, regarding evaluation frameworks, while SolidityScore effectively measures semantic consistency, static metrics inherently fail to reflect runtime behavior. Future benchmarks should integrate dynamic analysis techniques (e.g., fuzzing and symbolic execution) to detect high-risk vulnerabilities (e.g., reentrancy). At the same time, establishing quantitative standards for gas efficiency is essential to shift the  evaluation focus  from mere syntactic correctness to execution safety and economic optimization.

Second, in terms of optimization methodologies, currently isolated approaches (e.g., RAG and SFT) each face inherent limitations, necessitating the development of hybrid architectures. Future research should explore new paradigms (e.g., retrieval-augmented fine-tuning) to examine how dynamic knowledge bases can effectively guide model parameter updates. Furthermore, rigorous comparisons between full-parameter fine-tuning and parameter-efficient methods are needed to identify the optimal trade-off between computational cost and domain adaptation performance.

Finally, regarding models and data foundations, it is crucial to investigate the scaling laws of repository-level Solidity code generation capabilities across different model sizes. Concurrently, larger and more diverse datasets covering complex logic and edge cases should be constructed. These datasets should be continuously updated to keep pace with the rapid evolution of Solidity language features and security standards, thereby ensuring the temporal relevance of model capabilities. Advancing these directions will significantly enhance the performance of LLMs in repository-level Solidity smart contract generation, providing reliable support for the blockchain ecosystem and offering important insights for other domain-specific code generation tasks.

\subsection{Threats to Validity}
Internal validity is primarily challenged by two factors. First, model performance is heavily sensitive to prompt template design. Although we standardized prompts using structured templates (e.g., CoT and SCoT) to ensure consistency, alternative prompting strategies could yield different results. Second, the implementation of our experimental scripts poses another potential threat. To address this, all scripts underwent rigorous verification and testing. Additionally, we utilized models from the official Hugging Face Hub to further ensure implementation correctness and reproducibility.

External validity is primarily concerned with the generalizability of our findings. First, potential dataset bias is a key consideration. Although we curated data from authoritative and representative platforms (e.g., Synthetix and OpenZeppelin) using rigorous cleaning protocols, SolidiBench may not encompass all rare syntactic features or edge-case business logic. Consequently, the applicability of our conclusions to smart contracts written in other languages (e.g., Vyper and Move) requires further validation. Second, the selection of LLMs constitutes another external threat. This work focused on three 7B-parameter open-source models (DeepSeek-Coder, CodeLlama, and Qwen2.5-Coder). Therefore, our findings may not directly extrapolate to larger-scale or proprietary models. Future work should empirically evaluate a broader range of LLMs on our benchmark. Finally, computational constraints precluded an exhaustive hyperparameter search. While we enforced unified training configurations to ensure fair comparisons, this uniform approach may not have unlocked the optimal performance of each model architecture. Nevertheless, given Solidity's dominance in the Ethereum ecosystem and the practical prevalence of 7B-scale models for private deployment, our findings retain significant reference value for the current industry landscape.

Construct validity is primarily affected by the evaluation metrics used. Consistent with previous general-purpose code generation studies, we employed BLEU to evaluate the performance of the studied LLMs on Solidity code generation.  In view of the limitations of BLEU, we introduced a novel metric to comprehensively evaluate model performance across different paradigms.  Furthermore, we assessed the quality of generated code using SCR and CSR and analyzed the causes of compilation failures  to understand the practical limitations of  the Solidity code generated by the studied LLMs.


\section{Conclusion}
\label{chap:chap8}
 This paper presents a systematic empirical study on repository-level Solidity smart contract generation using LLMs across different adaptation paradigms. To address the lack of datasets for repository-level Solidity code generation, we construct a specialized dataset, SolidiBench, consisting of 5,470 natural language-Solidity code pairs. Using this dataset, we empirically evaluate the performance of three LLMs, specifically Qwen2.5-Coder, DeepSeek-Coder, and CodeLlama, under multiple paradigms. These paradigms include zero-shot, CoT, ICL, RAG, and SFT. Furthermore, to overcome the limitations of traditional metrics such as BLEU and the domain shift issues inherent in general-purpose CodeBERTScore, we introduce a novel metric called SolidityScore to more accurately assess the quality of generated Solidity code .
 
 Our findings reveal that while general-purpose models perform poorly in zero-shot settings, CoT significantly improves logical coherence, and RAG effectively incorporates domain knowledge. Most notably, SFT leads to the most substantial performance breakthrough, achieving a qualitative leap in semantic accuracy by internalizing domain-specific constraints of Solidity. We conclude that a pipeline combining high-quality domain data with SFT represents the optimal pathway for building highly reliable smart contract generation models. This study not only establishes a rigorous performance benchmark but also provides a methodological roadmap for repository-level Solidity code generation.

\normalem
\bibliographystyle{ACM-Reference-Format}
\bibliography{main}

\end{document}